\documentclass[aip,pof,preprint,amsmath,amssymb]{revtex4-1}

\usepackage{graphicx}
\usepackage{dcolumn}
\usepackage{multirow}
\usepackage{xcolor}
\usepackage{bm}

\newcommand{\be}{\begin{equation}}
\newcommand{\ee}{\end{equation}}
\newcommand{\bea}{\begin{eqnarray}}
\newcommand{\eea}{\end{eqnarray}}

\begin{document}

\title{Large-eddy simulation of Rayleigh-B\'{e}nard convection at extreme Rayleigh numbers}

\author{Roshan Samuel}
\email[]{roshanj@iitk.ac.in}
\affiliation{Department of Mechanical Engineering, Indian Institute of Technology, Kanpur 208016, India}

\author{Ravi Samtaney}
\email[]{ravi.samtaney@kaust.edu.sa}
\affiliation{Mechanical Engineering Program, Physical Science and Engineering Division, King Abdullah University of Science and Technology, Thuwal 23955, Saudi Arabia}

\author{Mahendra K. Verma}
\email[]{mkv@iitk.ac.in}
\affiliation{Department of Physics, Indian Institute of Technology, Kanpur 208016, India}

\date{\today}

\begin{abstract}
We adopt the stretched spiral vortex sub-grid model for large-eddy simulation (LES) of turbulent convection at extreme Rayleigh numbers.
We simulate Rayleigh-B\'{e}nard convection (RBC) for Rayleigh numbers ranging from \(10^6\) to \(10^{15}\) and for Prandtl numbers 0.768 and 1.
We choose a box of dimensions 1:1:10 to reduce computational cost.
Our LES yields Nusselt and Reynolds numbers that are in good agreement with the direct-numerical simulation (DNS) results of \citet{Iyer:PNAS2020},
albeit with a smaller grid size and at significantly reduced computational expense.
For example, in our simulations at \(Ra = 10^{13}\), we use grids that are 1/120 times the grid-resolution as that of the DNS \citep{Iyer:PNAS2020}.
The Reynolds numbers in our simulations span 3 orders of magnitude from 1,000 to 1,700,000.
Consistent with the literature, we obtain scaling relations for Nusselt and Reynolds numbers as \(Nu \sim Ra^{0.317}\) and \(Re \sim Ra^{0.497}\).
We also perform LES of RBC with periodic side-walls, for which we obtain the corresponding scaling exponents as \(0.343\) and \(0.477\) respectively.
Our LES is a promising tool to push simulations of thermal convection to extreme Rayleigh numbers, and hence 
enable us to test the transition to ultimate convection regime.
\end{abstract}


\maketitle


\section{Introduction}
\label{sec:intro}

Convective flows are frequently encountered in natural processes like atmospheric and planetary flows,
as well as in technological applications like electronic and industrial cooling systems.
Despite this ubiquity, turbulent convection at very high Rayleigh numbers of the order \(10^{15}\) is still poorly understood.
Unfortunately, due to the extreme complexity of the flow, direct numerical simulations do not reach such high Rayleigh numbers, except for the recent results of \citet{Iyer:PNAS2020}.
Consequently, we need robust subgrid-scale (SGS) models for simulating convection at very high Rayleigh numbers.

Rayleigh-B\'{e}nard convection (RBC) is an idealized setup used to study thermal convection.
Here, a layer of fluid is confined between two plates which are heated at the bottom and cooled at the top \cite{Getling:book, Verma:book:BDF}.
This causes the hot fluid to rise due to buoyancy while the cold fluid falls.
The two governing parameters of RBC are Rayleigh number (\(Ra\)), which is the ratio of buoyancy over dissipation,
and Prandtl number (\(Pr\)), which is the ratio of kinematic and thermal diffusivities.
For chosen values of \(Ra\) and \(Pr\), the Rayleigh B\'{e}nard flow is characterized by two dimensionless global quantities - Reynolds number (\(Re\)) and Nusselt number (\(Nu\)).
\(Re\) is a measure of the intensity of turbulence, whereas \(Nu\) quantifies the level of heat transfer from the top and bottom plates.
The scaling of these global quantities with \(Ra\) and \(Pr\) is a subject of ongoing research \cite{Castaing:JFM1989, Siggia:ARFM1994, Kadanoff:PT2001, Ahlers:RMP2009, Bhattacharya:PF2021}.

We know that in RBC, below a critical Rayleigh number (\(Ra_c\)), convection is suppressed by the viscosity of the fluid, and thermal conduction is the dominant mode of heat-transfer.
Above \(Ra_c\), convective flow is generated between the plates, leading to the formation of boundary layers at the top and bottom plates.
With increasing \(Ra\), the strength of flow in the bulk of the fluid increases and the boundary layers become thinner \cite{Grossmann:JFM2000, Bhattacharya:PF2021}.
\citet{Kraichnan:PF1962Convection} suggested that at very high \(Ra\), the boundary layers also become turbulent,
leading to increased heat transfer rates and a revised scaling for \(Nu\), called the ``ultimate regime''.
The possibility of existence of such an ultimate convection regime continues to motivate experiments of RBC at very high \(Ra\).

Here we focus on some of the numerical experiments performed at moderate to high Rayleigh numbers.
\citet{Verzicco:JFM2003} performed DNS of RBC at \(Ra\) up to \(2 \times 10^{11}\) and \(Pr = 0.7\) in a cylindrical domain of aspect ratio \(\Gamma = 1/2\)
to demonstrate the presence of a large-scale circulation (LSC) flow.
To investigate the dependence of aspect ratio on the formation of LSC, \citet{BailonCuba:JFM2010} performed DNS for \(0.5 \leq \Gamma \leq 12\) at the same \(Pr\),
but for a slightly reduced range of \(10^7 \leq Ra \leq 10^9\).
More recently, \citet{Foroozani:JFM2021} investigated the influence of different thermal boundary conditions
on the characteristics of turbulent convection at \(Pr = 0.7\) and 0.033, and \(Ra = 10^7\) and \(10^8\).
To the best of our knowledge, the highest Rayleigh numbers attained in direct numerical simulations of RBC were reported by \citet{Iyer:PNAS2020}.
They performed simulations in a slender cylindrical column of aspect-ratio 1/10 for \(10^8 \leq Ra \leq 10^{15}\) and \(Pr = 1\).

Note that these numerical studies and many others \cite{Shishkina:JFM2009, Stevens:JFM2010, Kooij:CAF2018} are performed in a cylindrical domain.
Similarly, there are also a number of experimental results for convection inside cylinders at very high Rayleigh numbers \cite{Niemela:JFM2003, Urban:PRL2011}.
However, various numerical studies have also been reported for convection in a box \cite{Kumar:PRE2014, Pandey:PRE2016}.
For instance, \citet{Wagner:PF2013} reported a detailed study on the dependence of aspect ratio on thermal convection in a box
for \(Pr = 0.786\) and for \(Ra\) ranging between \(10^5\) and \(10^9\).
\citet{Chong:PRF2018} further extended this study by performing DNS for a wider range of Prandtl numbers from 0.1 to 40 and a finer range of aspect ratios, but for a fixed Rayleigh number of \(10^8\).
\citet{Verma:NJP2017} performed pseudo-spectral simulations of RBC at \(Ra = 1.1 \times 10^{11}\)
and \(Pr = 1.1\) on a \(4096^3\) grid and showed that the kinetic energy energy spectrum shows Kolmogorov scaling at Prandtl numbers close to unity.
A detailed analysis on the importance of container shape on the dynamics of Rayleigh B\'{e}nard flow was recently presented by \citet{Shishkina:PRF2021}.

Direct numerical simulations of convection are usually limited to moderate \(Ra\) due to the increase in Reynolds number with Rayleigh number.
\(Re\) varies as \(Re \sim Ra^{\alpha}\), where \(\alpha\) has been shown to lie between 0.38 and 0.5 at high Rayleigh numbers \cite{Verma:book:BDF}.
With this increase in \(Re\), finer grids are necessary to resolve all the scales of the flow.
In this context, large-eddy simulation (LES) offers an attractive alternative for simulating RBC at very high Rayleigh numbers.
LES relies on robust models to approximate the effects of subgrid scale (SGS) dynamics on the resolved flow.
One of the earliest attempts to use subgrid modelling to simulate thermal convection was by \citet{Eidson:JFM1985}.
\citet{Kimmel:PF2000} devised an SGS model which estimates the effect of SGS eddies by expanding the resolved velocity and temperature fields.
This model was used to perform LES of RBC at \(Ra = 10^8\).
\citet{Foroozani:PRE2017} performed LES of turbulent thermal convection at \(Pr = 0.7\) and \(Ra\) of \(10^6\) and \(10^8\)
using the Lagrangian dynamic eddy-viscosity model \cite{Meneveau:JFM1996}.
More recently, \citet{Sondak:MRC2021} used the variational multiscale formulation to perform LES of RBC in a cylinder of aspect-ratio 1/4,
and attained \(Ra\) of up to \(10^{14}\) with a fluid of \(Pr = 7\).

One goal of LES is to accurately predict the momentum transfer due to the turbulent eddies at the subgrid scale.
\citet{Vashishtha:PRE2018} used the results from renormalization group (RG) theory \cite{McComb:book:HIT, Verma:PR2004},
to model this additional diffusion as a renormalized viscosity \cite{Vashishtha:PF2019}, and simulated RBC at \(Ra\) up to \(10^8\).
These simulations were performed using the pseudo-spectral solver, TARANG \cite{Verma:Pramana2013tarang}, with free-slip boundary conditions.

The stretched-spiral vortex (SSV) model for LES was originally developed by \citet{Misra:PF1997} for pseudo-spectral solvers by modelling the
fine structures of turbulence at the subgrid-scale (SGS) as strained spiral vortices \cite{Lundgren:PF1982}.
\citet{Kosovic:PF2002} used the SSV model (and also a nonlinear SGS model not discussed here) with a spectral solver to perform LES of compressible decaying turbulence in a periodic box.
Meanwhile \citet{Voelkl:PF2000} repurposed the model to compute subgrid stress in the physical space by using second-order structure-functions.
Moreover, \citet{Chung:JFM2009} augmented the SSV LES with a near-wall subgrid-scale model to simulate turbulent channel flow.
The wall-modelled LES has since been used to simulate several canonical flows like turbulent flat-plate boundary layer \cite{Inoue:JFM2011, Cheng:CF2014, Cheng:JFM2015}, 
flow past airfoils \cite{Gao:JFM2019}, Taylor-Couette flow \cite{Cheng:JFM2020}, flow in a periodically constricted channel \cite{Gao:JFM2020} and most recently,
plane Couette flow \cite{Cheng:JFM2022}.

In all the implementations of SSV model for LES discussed up till now, the subgrid closure is utilised only for the momentum equation,
that is, they are examples of hydrodynamic turbulence.
However, the model has also been extended to compute the subgrid flux of a passive scalar by \citet{Pullin:PF2000}.
\citet{Chung:JFM2010} used this SGS scalar flux closure for passive scalars to perform LES of stationary buoyancy-driven turbulent mixing of an active scalar in a triply periodic domain.
Subsequently \citet{Chung:JAS2014} extended the model by incorporating effects of buoyancy on the SGS turbulent kinetic energy and eddy length scale, under both stable and unstable stratification.
The present work is the first application of the SSV model for LES of the canonical Rayleigh-B\'{e}nard convection at very high Rayleigh numbers.

In the following sections, we present and analyze the results from this novel implementation of LES for RBC.
We first outline the governing equations and briefly describe the numerical method and LES model in Section~\ref{sec:govern_eqns}.
In Section~\ref{sec:les_dns}, we verify the subgrid-scale model against DNS results of \citet{Wagner:PF2013} for RBC in a box.
The LES is shown to clearly reproduce the effect of aspect ratio on thermal convection in a box,
as well as match the distribution of kinetic energy across 4 modes for \(Ra\) ranging from \(10^6\) to \(10^9\).
Moreover, the turbulent fluctuations of velocity and temperature are also captured very well by the LES model.
In Section~\ref{sec:les_rbc}, we present our results on RBC at very high \(Ra\) in a thin columnar box 
and reproduce the scaling relations obtained by \citet{Iyer:PNAS2020}.
The effect of side-walls are then eliminated by repeating a subset of the numerical experiments from Section~\ref{sec:les_rbc} but with periodic boundary conditions in the two horizontal directions.
We analyse the modifications in the scaling of \(Nu\) and \(Re\) triggered by removing the adiabatic side-walls in Section~\ref{sec:rbc_per}.
Finally we summarize and conclude our findings in Section~\ref{sec:conclusions}.


\section{Governing Equations and LES Model}
\label{sec:govern_eqns}

Here we describe the governing equations and the configuration of domains used in our numerical experiments.
We present a brief overview of the numerical methods, while a detailed description has been relegated to Appendix~\ref{app:num_meth}.
There is an extant literature on the stretched-spiral vortex model for large-eddy simulations of hydrodynamic turbulence \cite{Chung:JFM2009, Inoue:JFM2011, Gao:JFM2019}.
Here, we provide a brief overview of the model.
We will focus on the derivation of the subgrid temperature-flux, which accounts for the turbulent transport of heat in our LES.

\begin{figure}
    \centering
    \includegraphics[width=0.6\textwidth]{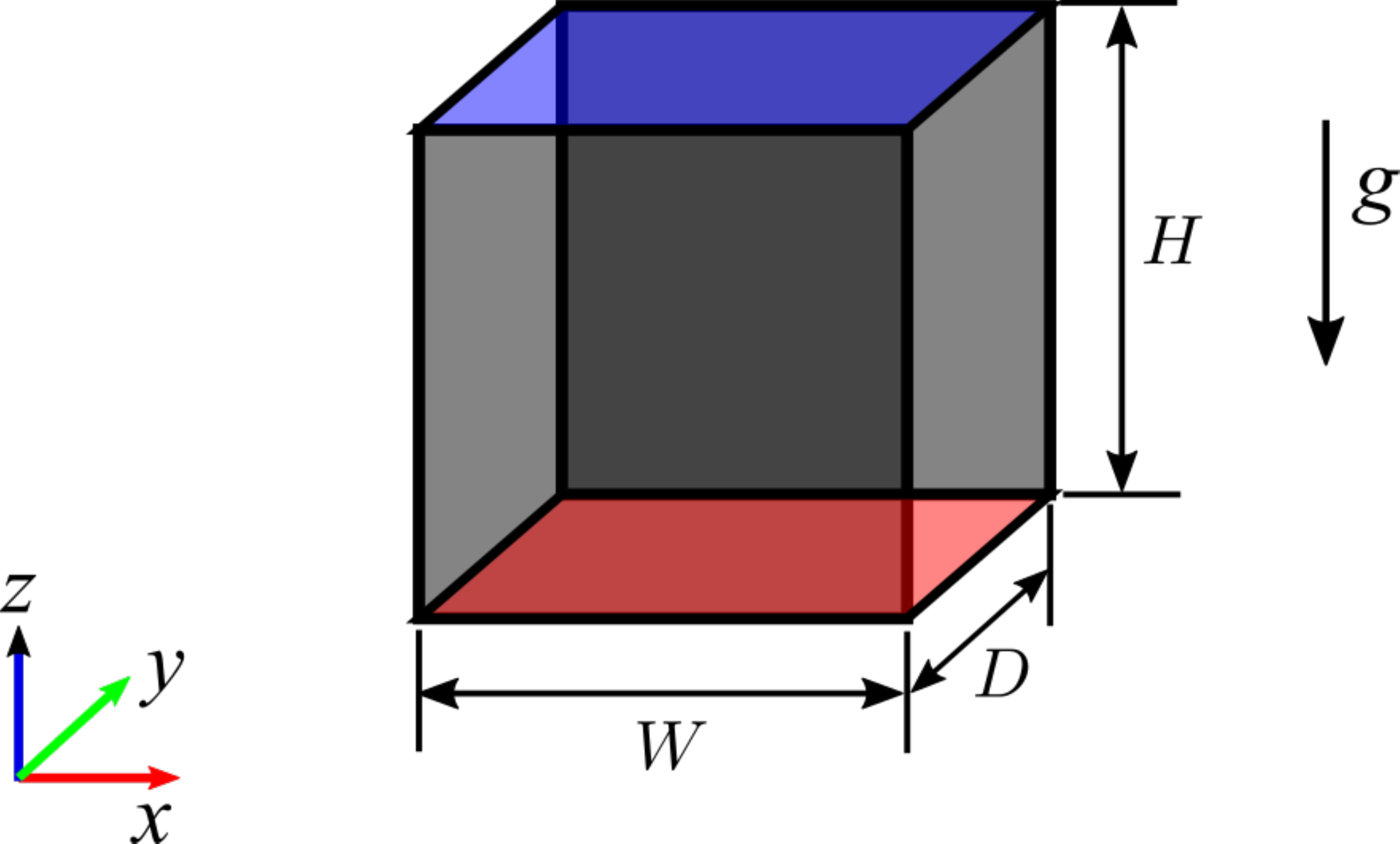}
    \caption{\label{fig:domains}
        Sketch of the computational domain for the LES of RBC.
        The fluid is confined between the heated bottom plate (red) and cooled top plate (blue).
        We report results for both non-periodic (adiabatic, no-slip), as well as periodic side-walls.
        The width (\(W\)), depth (\(D\)) and height (\(H\)) are adjusted as described in the text.
        For the cases described in Section~\ref{sec:les_dns}, \(W = H = 1\) and the aspect ratio, defined as \(\Gamma = D/H\), is varied by changing \(D\) only.
        In Section~\ref{sec:les_rbc} and \ref{sec:rbc_per}, \(H = 1\), \(W = D\), and the aspect ratio \(\Gamma = D/H = W/H = 0.1\) for all the cases.
            }
\end{figure}


\subsection{Governing Equations}
\label{sec:nse}

We solve the incompressible Navier-Stokes equation along with the temperature equation to compute the velocity field \(\mathbf{u}\) and the temperature field \(T\) respectively.
The two equations are coupled by the Oberbeck-Boussinesq approximation, wherein density variations are ignored everywhere except to account for the effects of buoyancy,
which manifests as a forcing term in the momentum equation.
Non-Boussinesq effects are normally expected for \(Ra > 10^{15}\), which is beyond the limit of the present study \cite{Niemela:Nature2000}.
Thus, the full set of equations are
\be
    \frac{\partial \mathbf{u}}{\partial t} + \mathbf{u}.\nabla\mathbf{u} = -\frac{1}{\rho_0}\nabla p + \alpha g T \hat{z} + \nu \nabla^2 \mathbf{u},
\ee
\be
    \frac{\partial T}{\partial t} + \mathbf{u}.\nabla T = \kappa \nabla^2 T,
\ee
\be
    \nabla.\mathbf{u} = 0.
\ee
Here, \(p\) is the pressure field, \(\alpha\) is the thermal expansion coefficient,
\(g\) is the gravitational acceleration, \(\nu\) and \(\kappa\) are the kinematic viscosity and thermal diffusivity respectively,
and \(\rho_0\) is the constant value of density within the Boussinesq approximation.
The equations are non-dimensionalized by the free-fall velocity \(u_f = \sqrt{\alpha g \Delta H}\), imposed temperature difference \(\Delta\), and domain height \(H\),
so that we get
\be
    \label{eq:eqn_vel}
    \frac{\partial \mathbf{u}}{\partial t} + \mathbf{u}.\nabla\mathbf{u} = -\nabla p + T \hat{z} + \sqrt{\frac{Pr}{Ra}} \nabla^2 \mathbf{u},
\ee
\be
    \label{eq:eqn_temp}
    \frac{\partial T}{\partial t} + \mathbf{u}.\nabla T = \frac{1}{\sqrt{Ra Pr}} \nabla^2 T.
\ee
We now have the two dimensionless parameters, Rayleigh number, \(Ra = \alpha g \Delta H^3 / (\kappa \nu)\), and Prandtl number, \(Pr = \nu/\kappa\).
We perform our 3D simulations in a cubical domain of height \(H\), width \(W\), and depth \(D\) as sketched in Fig.~\ref{fig:domains}.
The aspect ratio of the domain is defined as \(\Gamma = D/H\).
The fluid inside the cell is heated at the bottom and cooled at the top by setting the dimensionless temperatures \(T = 0\) and 1 
at the top (blue) and bottom (red) plates respectively.
At the four faces of the domain facing the horizontal directions, we use either no-slip adiabatic walls (Section~\ref{sec:les_dns} and \ref{sec:les_rbc}), or periodic boundary conditions (Section~\ref{sec:rbc_per}).


\subsection{Sub-grid Model}
\label{sec:model}

In our LES, we solve (\ref{eq:eqn_vel}) and (\ref{eq:eqn_temp}) on a coarse grid which does not resolve all the scales of the flow.
Hence, the velocity, pressure and temperature field are implicitly filtered by the grid spacing,
and decomposed into resolved and unresolved parts as, \(u_i = \tilde{u}_i + u_i'\),
\(p = \tilde{p} + p'\) and \(T = \tilde{T} + T'\) respectively.
We then obtain the filtered governing equations on the LES grid as
\be
    \label{eq:nse_filt}
    \frac{\partial \tilde{u}_i}{\partial t} + \tilde{u}_j \frac{\partial \tilde{u}_i}{\partial x_j} =
    -\frac{\partial \tilde{p}}{\partial x_i} + \tilde{T} \delta_{i3} + \sqrt{\frac{Pr}{Ra}} \frac{\partial^2 \tilde{u}_i}{\partial x_j^2} - \frac{\partial \tau_{ij}}{\partial x_j},
\ee
\be
    \label{eq:temp_filt}
    \frac{\partial \tilde{T}}{\partial t} + \tilde{u}_j \frac{\partial \tilde{T}}{\partial x_j} =  \sqrt{\frac{1}{Ra Pr}} \frac{\partial^2 \tilde{T}}{\partial x_j^2} - \frac{\partial \sigma_j}{\partial x_j}.
\ee
Here, \(\tau_{ij} = \widetilde{u_i u_j} - \tilde{u}_i \tilde{u}_j\) is the subgrid stress tensor,
and \(\sigma_j = \widetilde{T u_j} - \tilde{T} \tilde{u}_j\) is the subgrid scalar flux.
Both the terms account for the effects of unresolved subgrid components of velocity and temperature respectively.
The stretched-vortex model developed by \citet{Misra:PF1997} computes \(\tau_{ij}\) by assuming that the subgrid flow is approximated
by a stretched spiral vortex \cite{Lundgren:PF1982}.
Here the subgrid stress tensor is represented as
\be
    \label{eq:les_model}
    \tau_{ij} = (\delta_{ij} - e^v_i e^v_j)K,
\ee
where \(K\) is the subgrid kinetic energy and \(\mathbf{e}^v\) is the unit vector aligned along the axis of the subgrid vortex.
We note that
\be
    \label{eq:sgv_K}
    K = \int_{k_c}^{\infty}E(k) dk,
\ee
where \(E(k)\) is the subgrid energy spectrum, and \(k_c\) is the cutoff wavenumber which is defined by the resolved grid spacing \(h_c\) as \(k_c = \pi/h_c\).
Here, \(h_c = (h_x h_y h_z)^{1/3}\) is also the length-scale of subgrid motions, which will be used subsequently in the derivation of subgrid scalar flux.
\citet{Lundgren:PF1982} derived \(E(k)\) for the stretched subgrid vortex as
\be
    \label{eq:sgv_Ek}
    E(k) = \mathcal{K}_0 \epsilon^{2/3} k^{-5/3} \exp{\left[\frac{-2 \nu k^2}{3|\tilde{a}|}\right]}.
\ee
Here \(\mathcal{K}_0\) is a prefactor, \(\epsilon\) is the dissipation rate, \(\nu\) is kinematic viscosity,
and \(\tilde{a}\) is a measure of the stretching experienced by the subgrid vortex due to the resolved strain-rate tensor, and it is written as
\be
    \tilde{a} = e^v_i e^v_j \tilde{S}_{ij}, \qquad \qquad \tilde{S}_{ij} = \frac{1}{2}\left(\frac{\partial \tilde{u}_i}{\partial x_j} + \frac{\partial \tilde{u}_j}{\partial x_i}\right).
\ee
Substituting (\ref{eq:sgv_Ek}) into (\ref{eq:sgv_K}) and integrating from \(k_c\) to \(\infty\) yields the following expression for \(K\):
\be
    \label{eq:les_K}
    K = \frac{\mathcal{K}_0 \epsilon^{2/3}}{2} \left(\frac{2 \nu}{3|\tilde{a}|}\right)^{1/3} \Gamma{\left[-\frac{1}{3}, \frac{2 \nu k_c^2}{3|\tilde{a}|}\right]},
\ee
where \(\Gamma[a, b]\) is an incomplete gamma function.
The prefactor \(\mathcal{K}_0\) is calculated using local structure functions as described by \citet{Voelkl:PF2000}.
The module for computing \(K\) as given by (\ref{eq:les_K}) using approximations for the gamma function is provided in an open-source code from Pullin and coworkers.
After obtaining \(K\), the only remaining unknown in (\ref{eq:les_model}) is \(\mathbf{e}^v\).
Different variations of the LES model based on the choice of \(\mathbf{e}^v\) are explored by \citet{Misra:PF1997}.
In our simulations, we choose \(\mathbf{e}^v\) to be aligned along the most extensive eigenvector of \(\tilde{S}_{ij}\).

The novel aspect of the present work lies in the use of SSV model to simulate RBC at very high Rayleigh numbers.
For this we compute the subgrid scalar flux \(\bm{\sigma}\) which closes the non-linear term of the temperature equation.
\citet{Pullin:PF2000} derived an expression for \(\sigma_j\) by first deriving the subgrid scalar flux in the reference frame attached to the subgrid vortex,
\be
    \label{eq:sigma_sg}
    \sigma_j' = -\frac{K_v t_{\mathrm{SG}}}{2}\left(\frac{\partial \tilde{T}}{\partial x_j'}\right), \qquad j = 1, 2.
\ee
Here, \(K_v\) is the kinetic energy contained in the subgrid vortex, and \(t_\mathrm{SG}\) is the typical eddy turnover time for the subgrid motions.
Note that \(x_j'\) is defined in the vortex frame of reference, and the scalar gradient in this frame is computed along the plane perpendicular to the vortex (spanned by \(e_1'\) and \(e_2'\)).
Using the transformation matrix derived by \citet{Misra:PF1997} for calculating \(\bm{\tau}\), (\ref{eq:sigma_sg}) can be transformed to the global coordinates, \(x_i\), to give
\be
    \label{eq:sigma_glob}
    \sigma_j = -\frac{K_v t_\mathrm{SG}}{2}(\delta_{jp} - e_j^v e_p^v)\frac{\partial \tilde{T}}{\partial x_p}.
\ee
It was deduced that the vortex energy \(K_v\) is identical to the subgrid kinetic energy, \(K\).
Moreover, the subgrid eddy turnover time can be written as \(t_\mathrm{SG} = K/\epsilon_\mathrm{SG}\), where \(\epsilon_\mathrm{SG}\) is the subgrid energy dissipation rate \cite{Chung:JAS2014}.
Using a dimensionless constant \(\gamma\) of \(O(1)\), \(\epsilon_\mathrm{SG}\) is written in terms of \(K\) and cut-off length scale, \(h_c\), as
\be
    \epsilon_\mathrm{SG} = \frac{K^{3/2}}{\gamma h_c}.
\ee
We now have \(t_\mathrm{SG} = \gamma h_c/\sqrt{K}\), which can be substituted in (\ref{eq:sigma_glob}) to give the required closure for the filtered temperature equation, (\ref{eq:temp_filt}),
\be
    \bm{\sigma} = -\frac{\gamma \pi}{2 k_c}\sqrt{K}\left[\mathbf{I} - (\mathbf{e}^v \otimes \mathbf{e}^v)\right]\nabla \tilde{T}.
\ee
We already have the values of \(K\) and \(\mathbf{e}^v\) from the calculation of \(\bm{\tau}\) earlier.
Once we fix the value of \(\gamma\), we will be in a position to compute \(\bm{\sigma}\) from the resolved temperature gradient vector \(\nabla \tilde{T}\).
Using an argument similar to that used by \citet{Lilly:book_chapter} to obtain the Smagorinsky constant, \citet{Pullin:PF2000} estimated \(\gamma\) and found that it lies in the range of 0.89 to 1.3.
All realizations of the SSV model in literature use \(\gamma = 1.0\) for their calculations, and we retain this value in all our simulations.
More details of the derivation can be found in \citet{Pullin:PF2000} and \citet{Chung:JAS2014}.


\subsection{Numerical Method}
\label{sec:method}

We use the open-source finite-difference code, SARAS \cite{Samuel:JOSS2021}, for the simulations presented here.
As a general purpose partial differential equation solver, SARAS has been validated extensively for different types of fluid flow problems.
\citet{Verma:SN2020} compared its performance in DNS of decaying turbulence against the pseudo-spectral solver, TARANG \cite{Verma:Pramana2013tarang}.
Moreover, \citet{Bhattacharya:PF2021} extended the Grossman and Lohse's model \cite{Grossmann:PRL2001} for \(Re\) and \(Nu\) scaling in turbulent RBC using DNS results from SARAS.
More recently, SARAS was used to highlight the Prandtl number dependence of small scale quantities like
energy dissipation rates and structure functions in turbulent RBC \cite{Bhattacharya:PRF2021}.
All these results used the capacity of SARAS to perform DNS at high resolutions, albeit at Rayleigh numbers less than \(10^{10}\).
In the present work, we extend the capability of SARAS to even higher \(Ra\) using its newly implemented stretched-spiral vortex LES module.

We use a third-order low-storage Runge-Kutta scheme \cite{Spalart:JCP1991} to solve the governing equations.
Here the diffusion terms of both velocity and temperature are calculated semi-implicitly, whereas the remaining terms are treated explicitly.
Moreover, the non-linear terms are split into two according to the skew symmetric form described by \citet{Morinishi:JCP1998}.
SARAS uses the geometric multigrid method with V-cycles to solve the pressure Poisson equation.
The solver is parallelized by decomposing the computational domain along all the three axes.
A more detailed description of the numerical schemes and the implementation of the subgrid model is given in Appendix~\ref{app:num_meth}.


\section{Comparison of LES with DNS}
\label{sec:les_dns}

Presently, we verify the subgrid scalar flux computation of the stretched spiral vortex model by performing LES of thermal convection in a box.
\citet{Wagner:PF2013} performed direct numerical simulations of Rayleigh B\'{e}nard convection in a box for varying aspect-ratios in the range of \(1/10 \leq \Gamma \leq 1\).
We keep \(W = H = 1\) as constant, and \(\Gamma\) is varied by changing the value of depth \(D\) only (see Fig.~\ref{fig:domains}).
The DNS were performed at a constant Prandtl number of 0.786, and for Rayleigh numbers ranging from \(10^5\) to \(10^9\).
To verify our sub-grid model, we first perform LES at the same \(Pr\) and for \(10^6 \leq Ra \leq 10^9\) at a fixed \(\Gamma = 1/4\).
Subsequently, we also perform a separate set of LES at \(Ra = 10^7\), for \(\Gamma = 1/10, 1/4, 1/2\) and 1, in order to verify the temperature and velocity fluctuation profiles.

\begin{table}
    \caption{\label{table:wag_ar4_cases}
        Details of Rayleigh number (\(Ra\)) and grid sizes (\(\mathrm{N}_x \times \mathrm{N}_y \times \mathrm{N}_z\)) -- both for the LES results discussed in Section~\ref{sec:les_dns},
        and the DNS cases from \citet{Wagner:PF2013} against which the LES has been verified.
        The Nusselt number (\(Nu\)), Reynolds number (\(Re\)) and total simulation time (\(t_\mathrm{max}\)) in free-fall units are also listed.
        All the cases are performed in a domain with fixed \(\Gamma = 1/4\).
    }
    \begin{tabular}{lccccccr}
    \multirow{2}{*}{\(Ra\)}  &            \multicolumn{2}{c}{Grid}                      &   \multicolumn{2}{c}{\(Nu\)}  & \multicolumn{2}{c}{\(Re\)}          & \multirow{2}{*}{\(t_\mathrm{max}\)}   \\
                         & LES                          & DNS                           & LES               & DNS       & LES                   & DNS       &       \\
    \hline
    \(1 \times 10^{6}\)  & $32 \times 16 \times 32$     & $96 \times 32 \times 96$      & \(8.14\)          & ~7.99     & \(155.7\)             & ~153.5    & 1000  \\
    \(2 \times 10^{6}\)  & $32 \times 16 \times 32$     & $96 \times 32 \times 96$      & \(10.6\)          & 10.16     & \(227.0\)             & ~224.8    & 1000  \\
    \(3 \times 10^{6}\)  & $64 \times 32 \times 64$     & $192 \times 64 \times 192$    & \(10.9 \pm 1.9\)  & 10.42     & \(250.1 \pm 27.5\)    & ~248.7    & 1000  \\
    \(1 \times 10^{7}\)  & $64 \times 32 \times 64$     & $192 \times 64 \times 192$    & \(16.7 \pm 2.9\)  & 16.37     & \(469.0 \pm 48.0\)    & ~468.7    & 890   \\
    \(1 \times 10^{8}\)  & $128 \times 64 \times 128$   & $384 \times 128 \times 384$   & \(32.8 \pm 6.3\)  & 32.34     & \(1531 \pm 147\)      & 1553~     & 400   \\
    \(1 \times 10^{9}\)  & $256 \times 128 \times 256$  & $768 \times 256 \times 768$   & \(64.1 \pm 7.9\)  & 63.27     & \(5005 \pm 238\)      & 4995~     & 400   \\
    \hline
    \end{tabular}
\end{table}

While the reported DNS was performed on grids up to \(768 \times 256 \times 768\) points in size,
the maximum size of our LES grids is \(256 \times 128 \times 256\).
The SGS model confers a significant computational benefit by requiring grids that are 20-40 times smaller than those of the DNS cases in the comparisons presented here.
The grid-points are non-uniformly spaced so that the boundary layers near the no-slip walls on all sides of the domain are resolved.
We generate the non-uniform spacing using a tangent-hyperbolic function, and regulate the stretching parameters such that the cell aspect-ratio does not exceed 5 anywhere in the domain.

Here and in subsequent sections, the LES model uses 2-4 times fewer points in each direction as compared to DNS grids.
The subgrid scale estimation model employed by \citet{Kimmel:PF2000} yielded a similar saving in grid sizes.
Similarly, the LES of RBC performed by \citet{Vashishtha:PRE2018} also used an LES grid that was 4 times coarser as compared to the DNS grid.


\subsection{Temporal Variation of Global Quantities}
\label{sec:nu_re_ts}

\begin{figure}
    \centering
    \includegraphics[width=1\textwidth]{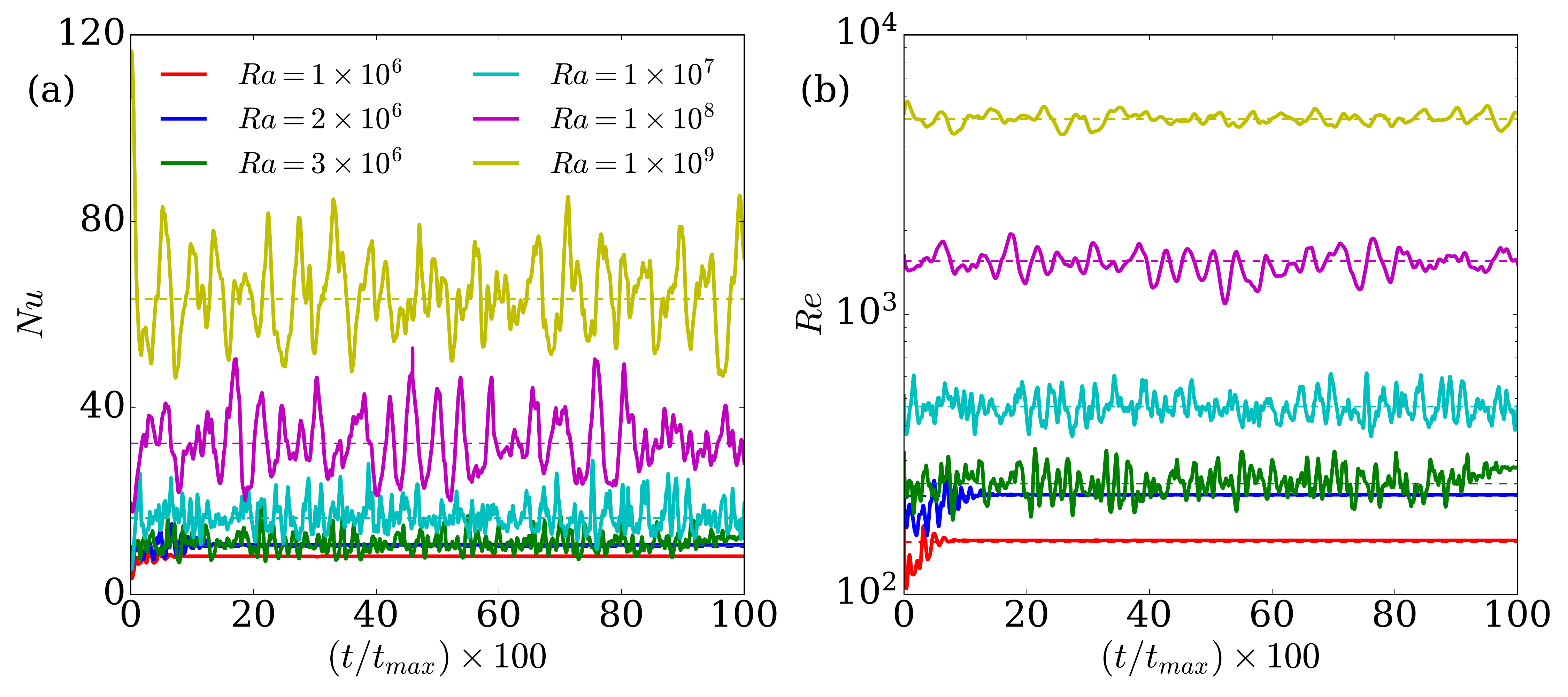}
    \caption{\label{fig:wagner_ts}
        Temporal variation of Nusselt and Reynolds numbers for LES of thermal convection at \(Pr = 0.786\)
        and \(10^6 \leq Ra \leq 10^9\) in a box of size \(1 \times 0.25 \times 1\) as described in the text.
        The time-spans have been normalized by the maximum time-units, \(t_{max}\), (given in Table~\ref{table:wag_ar4_cases})
        such that all runs have a common time-span from 0 to 100.
        The mean \(Nu\) and \(Re\) from the DNS result of \citet{Wagner:PF2013} are marked by thin dashed lines of the same color as the corresponding time-series data.
            }
\end{figure}

In Fig.~\ref{fig:wagner_ts}, we plot the temporal variation of Reynolds number, \(Re = \sqrt{Ra/Pr} \langle |\mathbf{u}| \rangle L \),
and Nusselt number, \(Nu = 1 + \sqrt{Ra Pr}\langle u_z T \rangle\), for the LES runs listed in Table~\ref{table:wag_ar4_cases}.
Note that both \(Nu\) and \(Re\) are calculated from the resolved velocity and temperature fields.
It is assumed that the SGS fluctuations of these fields will affect these global quantities negligibly.
The duration of the runs have been normalized by the maximum time-units elapsed (\(t_\mathrm{max}\)),
and mapped to a fixed interval \(t' \in [0, 100]\), so that all 6 simulations can be tracked over a normalized duration.
The values of \(t_\mathrm{max}\) for each case are listed in Table~\ref{table:wag_ar4_cases}.

For both the \(Nu\) and \(Re\) variations shown in Figs.~\ref{fig:wagner_ts}(a) and \ref{fig:wagner_ts}(b) respectively,
the corresponding DNS values from \citet{Wagner:PF2013} are plotted using thin dashed lines.
We note that an important aspect of thermal convection that the LES captures correctly is the onset of unsteady motion.
For the box with \(\Gamma = 1/4\), DNS results showed that the flow attained steady state for \(Ra \leq 2 \times 10^6\),
and was unsteady for all cases with \(Ra \geq 3 \times 10^6\).
Correspondingly, the time-series from the LES also shows that the convecting flow becomes unsteady for \(2 \times 10^6 \le Ra \leq 3 \times 10^6\).
The mean values for \(Re\) and \(Nu\) obtained from LES are also listed in Table~\ref{table:wag_ar4_cases}, and
we observe that the LES results match the DNS values reasonably well.


\subsection{Large-scale Structure of Convection}
\label{sec:wag_lss}

\begin{figure}
    \centering
    \includegraphics[width=1.0\textwidth]{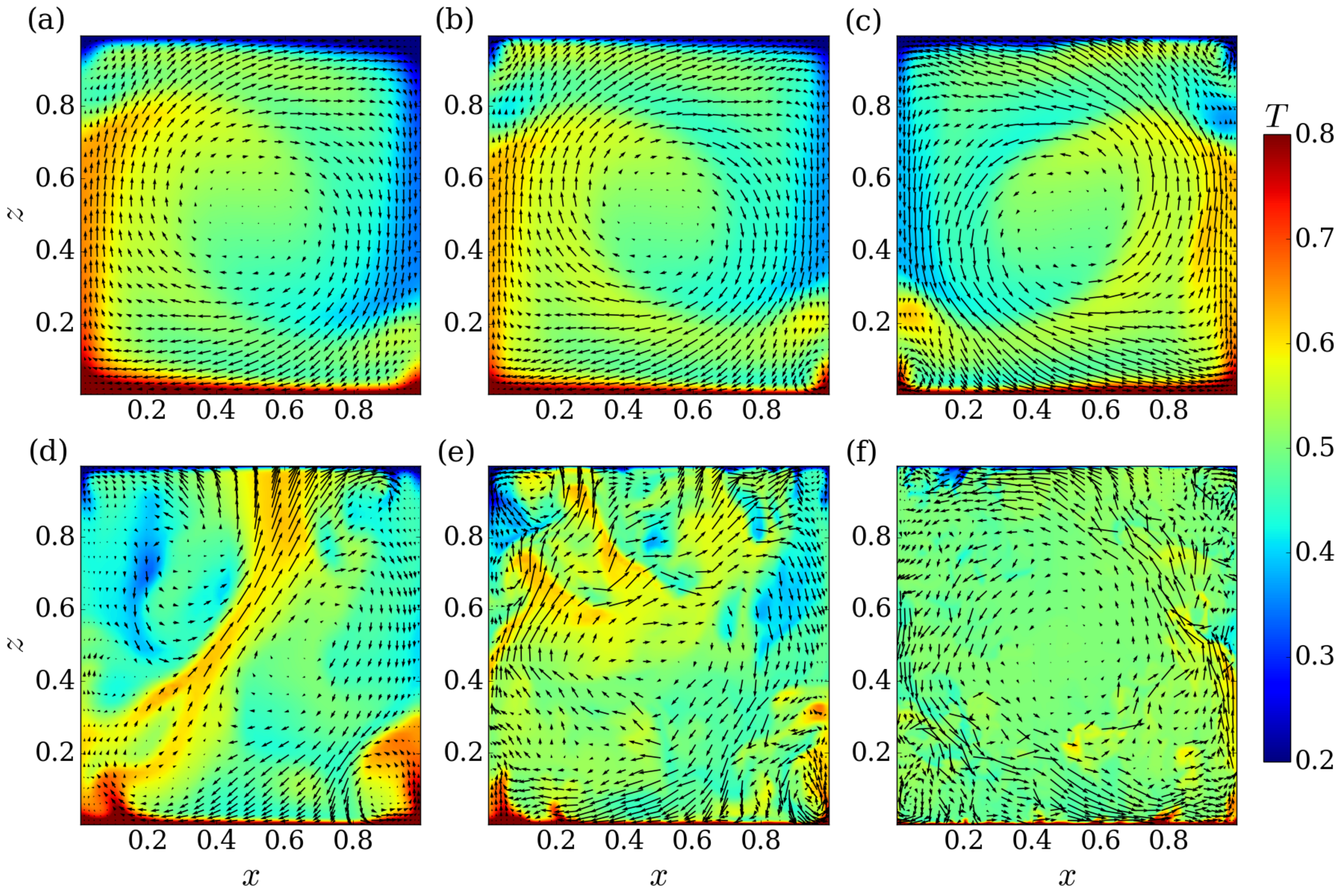}
    \caption{\label{fig:wagner_ff}
        Contour plots of instantaneous temperature field, \(T\), along with \((u_x, u_z)\) velocity vectors on the \(xz\)-plane at \(y = 0.125\) in a cubical box of dimensions \(1 \times 0.25 \times 1\).
        The snapshots are taken at (a) \(Ra = 1 \times 10^6\), (b) \(2 \times 10^6\), (c) \(3 \times 10^6\),
        (d) \(1 \times 10^7\), (e) \(1 \times 10^8\), and (f) \(1 \times 10^9\) for \(Pr = 0.786\).
        In the first three cases, the data is taken at \(t = 1000\), when (a) and (b) are in steady state, but (c) continues to vary with time and undergo flow reversals.
        The data for the last three cases have been taken at t = 700, 300 and 300 respectively.
            }
\end{figure}

In Fig.~\ref{fig:wagner_ff}, we plot the contours of instantaneous temperature fields on the mid \(y\)-plane.
To highlight the structure of flow in the cell, the plots are overlaid with the velocity vectors \((u_x, u_z)\) in the plane.
We observe that at low \(Ra\) of order \(10^6\), the flow is dominated by a well-defined central roll aligned along the y-axis (normal to the plane), as seen in the first row of Fig.~\ref{fig:wagner_ff}.
As mentioned before, the flow is steady in Figs.~\ref{fig:wagner_ff}(a) and \ref{fig:wagner_ff}(b), corresponding to \(Ra = 1 \times 10^6\) and \(2 \times 10^6\) respectively.
We also observe that the direction of the central roll is reversed in Fig.~\ref{fig:wagner_ff}(c), corresponding to \(Ra = 3 \times 10^6\).
This is because the unsteady flow at \(Ra = 3 \times 10^6\) undergoes multiple flow reversals, where the roll alternates between clockwise and anti-clockwise directions.
This is a comprehensively studied aspect of thermal convection and has also been established through low-dimensional models \cite{Chandra:PRE2011, Verma:book:BDF}.

For the higher range of \(10^7 \leq Ra \leq 10^9\) whose instantaneous temperature and velocity fields are shown in the second row of Fig.~\ref{fig:wagner_ff},
we observe that the flow is more chaotic with a somewhat less coherent roll in the bulk of the domain.
At \(Ra = 10^7\), shown in Fig.~\ref{fig:wagner_ff}(d), the snapshot of the flow field is taken at \(t = 700\) and shows the presence of multiple rolls.
These structures continually merge and split in the course of the unsteady flow.
Meanwhile, a large-scale circulation (LSC) flow is discernible for both \(Ra = 10^8\) and \(10^9\).
The LES also captures the increasing prominence of corner vortices as the \(Ra\) increases, as noted in previous DNS works \cite{Chandra:PRL2013, Wagner:PF2013}.
This observation is also reinforced by the variation in amplitudes of a few selected Fourier modes of the flow discussed in the next subsection.


\subsection{Modal Analysis}
\label{sec:wag_modal}

\begin{figure}
    \centering
    \includegraphics[width=1.0\textwidth]{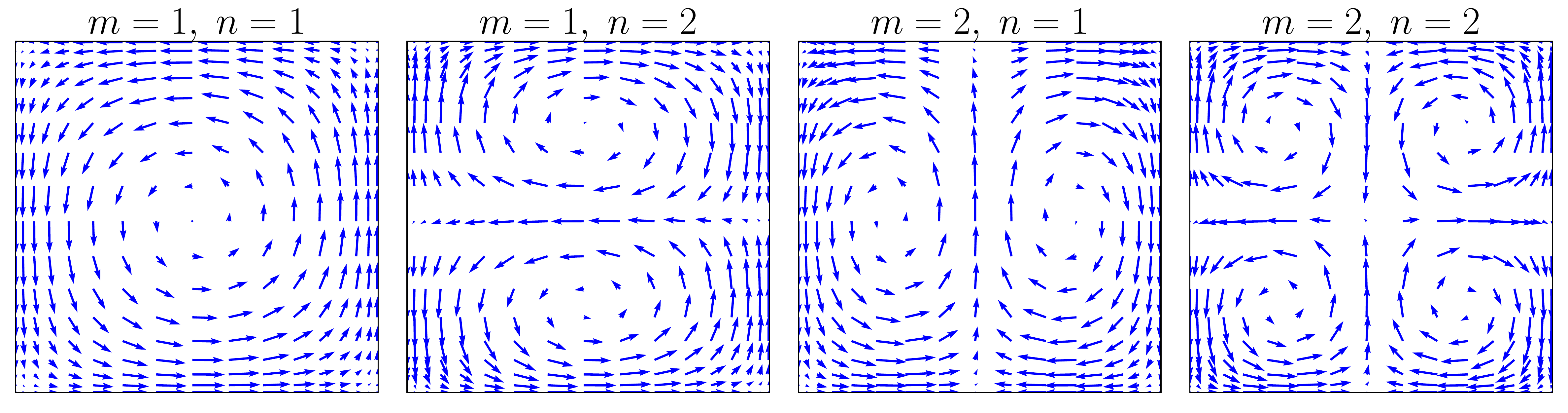}
    \caption{\label{fig:demo_modes}
        The four modes for \(u_x\) and \(u_z\) at the mid \(y\)-plane, defined according to (\ref{eq:wag_modes}), whose energy contributions are plotted in Fig.~\ref{fig:wagner_modes}.
            }
\end{figure}

We have seen that the structure of convective flow evolves through multiple patterns of rolls as we increase the Rayleigh number.
This intermittent behaviour of the large-scale flow in convection has been analysed in both 2D and 3D numerical studies \cite{Petschel:PRE2011, Mishra:JFM2011}.
\citet{Chandra:PRL2013} studied the mechanism responsible for these roll reversals, and highlighted the importance of vortex reconnections in restructuring the flow.
In their study, the flow structures were quantified by projecting the two components of velocity in a plane onto a sine-cosine Fourier basis \cite{Chandra:PRE2011}.
\citet{Ao:PF2020} also recently used the same decomposition along with proper orthogonal decomposition to extract coherent flow structures in RBC.
Although \citet{Chandra:PRL2013} performed the analysis using DNS results from 2D RBC, \citet{Wagner:PF2013} were able to meaningfully apply the same technique to
the velocity field on the mid-\(xz\)-plane of their 3D RBC results.
It was argued that the flow may be considered quasi two-dimensional in boxes with \(\Gamma \leq 1/3\),
due to the highly restricted depth which largely confines the mean flow to the \(xz\)-plane \cite{Xia:PRE2003}.
We now apply the same procedure to verify the fidelity of LES in predicting the average distribution of energy across a given set of modes.

For each of the 6 LES cases listed in Table~\ref{table:wag_ar4_cases}, we used the snapshots of the flow field along a vertical slice at half-depth
(one of which has been plotted for each case in Fig.~\ref{fig:wagner_ff}).
We consider only the two components of velocity in the plane, namely \(u_x\) and \(u_z\), and project them onto the Fourier basis
\bea
    v_x^{m, n} &=& 2 \sin{(m \pi x)} \cos{(n \pi z)}, \nonumber \\
    v_z^{m, n} &=&-2 \cos{(m \pi x)} \sin{(n \pi z)},
    \label{eq:wag_modes}
\eea
where \(m, n \in \{1, 2\}\).
Please note that the factor 2 appearing above in the basis functions ensure that the modes are normalized such that \(\iint v_j^{m, n} v_j^{m, n} dx dz = 1\) for any \((m, n)\) and \(j \in {x, z}\).
This gives us a set of four orthornormal modes - (1, 1), (1, 2), (2, 1), and (2, 2), whose resultant flows are plotted in Fig.~\ref{fig:demo_modes}.
The time-varying contributions from the two components of each mode are given by their scalar products with the corresponding velocity components in the plane
\be
    A_x^{m, n}(t) = \langle u_x(t) v_x^{m, n} \rangle_{x, z}, \qquad A_z^{m, n}(t) = \langle u_z(t) v_z^{m, n} \rangle_{x, z}.
\ee
Thus the contribution from a mode \((m, n)\) to the flow at time \(t\) is the \(\mathrm{L}_2\) norm of the two components, \(M^{m, n}(t) = \sqrt{A_x^{m, n}(t) + A_z^{m, n}(t)}\).
Finally we obtain the time-averaged contribution of each mode from the time-series.

\begin{figure}
    \centering
    \includegraphics[width=0.8\textwidth]{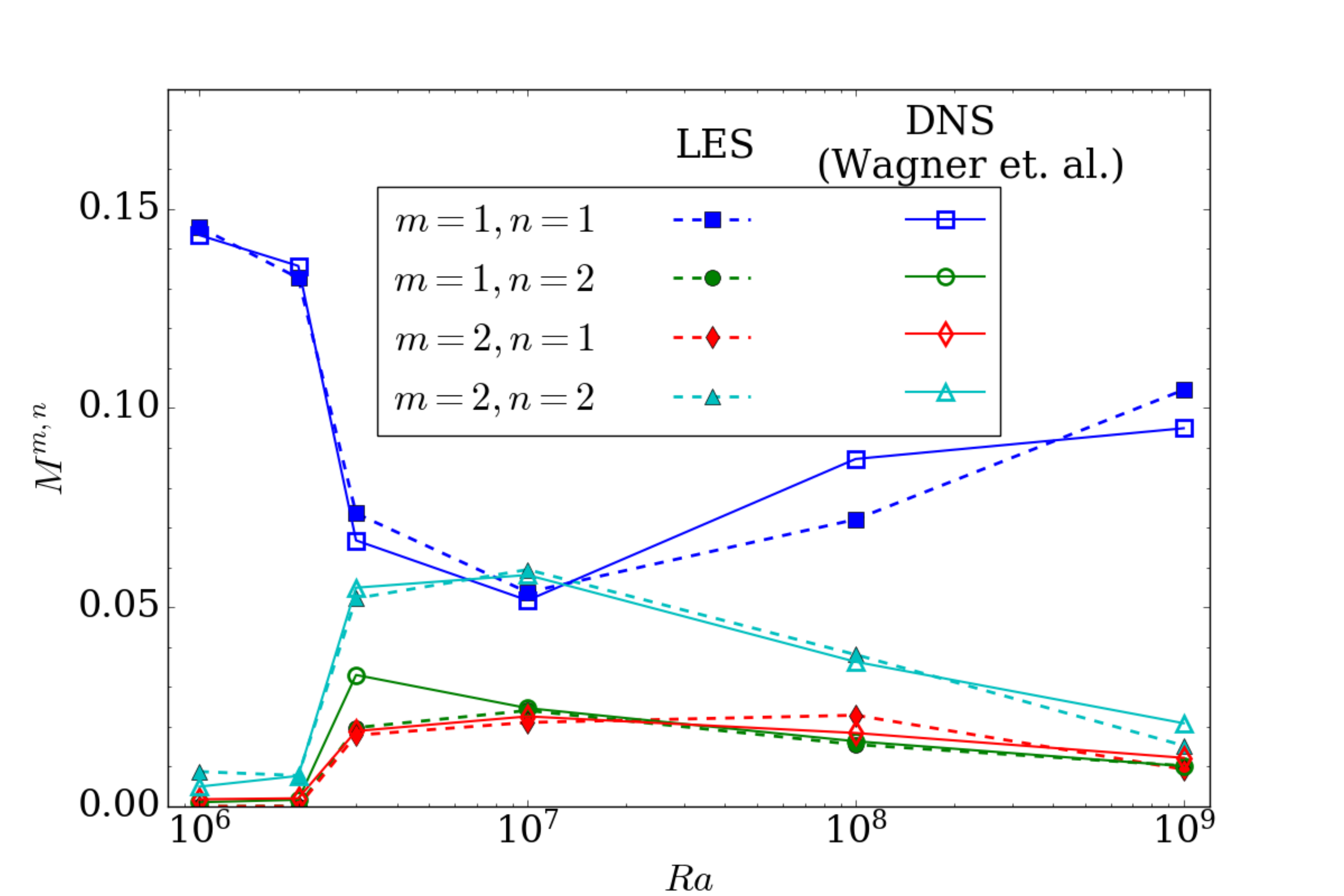}
    \caption{\label{fig:wagner_modes}
        Variation of energy contained in the 4 modes, (1, 1), (1, 2), (2, 1) and (2, 2), as defined by (\ref{eq:wag_modes}) and sketched in Fig.~\ref{fig:demo_modes}.
        The four modes are plotted with square, circle, diamond and triangle symbols respectively.
        While the LES data are plotted with filled symbols, the corresponding values from the DNS results of \citet{Wagner:PF2013} are marked by hollow symbols.
            }
\end{figure}

In Fig.~\ref{fig:wagner_modes}, we plot the variation of \(\langle M^{m, n} \rangle_t\) against \(Ra\) for each of the four modes.
The modal amplitude for the LES cases are plotted with filled symbols,
whereas the values from the modal analysis of DNS data from \citet{Wagner:PF2013} are plotted with corresponding unfilled symbols.
We observe that the LES predicts the dynamics of the flow with reasonable accuracy over the full range of Rayleigh numbers.
In the flow fields shown earlier in Fig.~\ref{fig:wagner_ff}, we had noted that the corner rolls increased in prominence with \(Ra\) up to \(10^7\),
where the 4-roll configuration is very much apparent.
This is reflected in the strength of mode (2, 2), which peaks at \(Ra\) of \(10^7\).
The appearance of this mode is also an indicator of flow reversals \cite{Chandra:PRL2013}, which was also observed in Fig.~\ref{fig:wagner_ff}.
The variation of (1, 1) mode, which denotes the central roll is also captured very well by the SGS model.
At low \(Ra\), this mode dominates by virtue of the steady central roll, whereas at high \(Ra\), the mode denotes the increasing strength of the LSC flow.


\subsection{Fluctuation profiles}
\label{sec:hf_profiles}

\begin{table}
    \caption{\label{table:wag_hf_cases}
        List of Rayleigh number \(Ra\), grid sizes, and aspect ratios \(\Gamma\) for the simulations performed to obtain the horizontal fluctuation profiles in Section~\ref{sec:hf_profiles}.}
    \begin{tabular}{lccc}
    \(Ra\)               & LES Grid                     & DNS Grid                      & \(\Gamma\)    \\
    \hline
    \(1 \times 10^{7}\)  & $64 \times 16 \times 64$     & $192 \times 64 \times 192$    & \(1/10\)      \\
    \(1 \times 10^{7}\)  & $64 \times 32 \times 64$     & $192 \times 64 \times 192$    & \(1/4\)       \\
    \(1 \times 10^{7}\)  & $64 \times 32 \times 64$     & $192 \times 128 \times 192$   & \(1/2\)       \\
    \(1 \times 10^{7}\)  & $64 \times 64 \times 64$     & $192 \times 192 \times 192$   & \(1\)         \\
    \hline
    \end{tabular}
\end{table}

\begin{figure}
    \centering
    \includegraphics[width=1.0\textwidth]{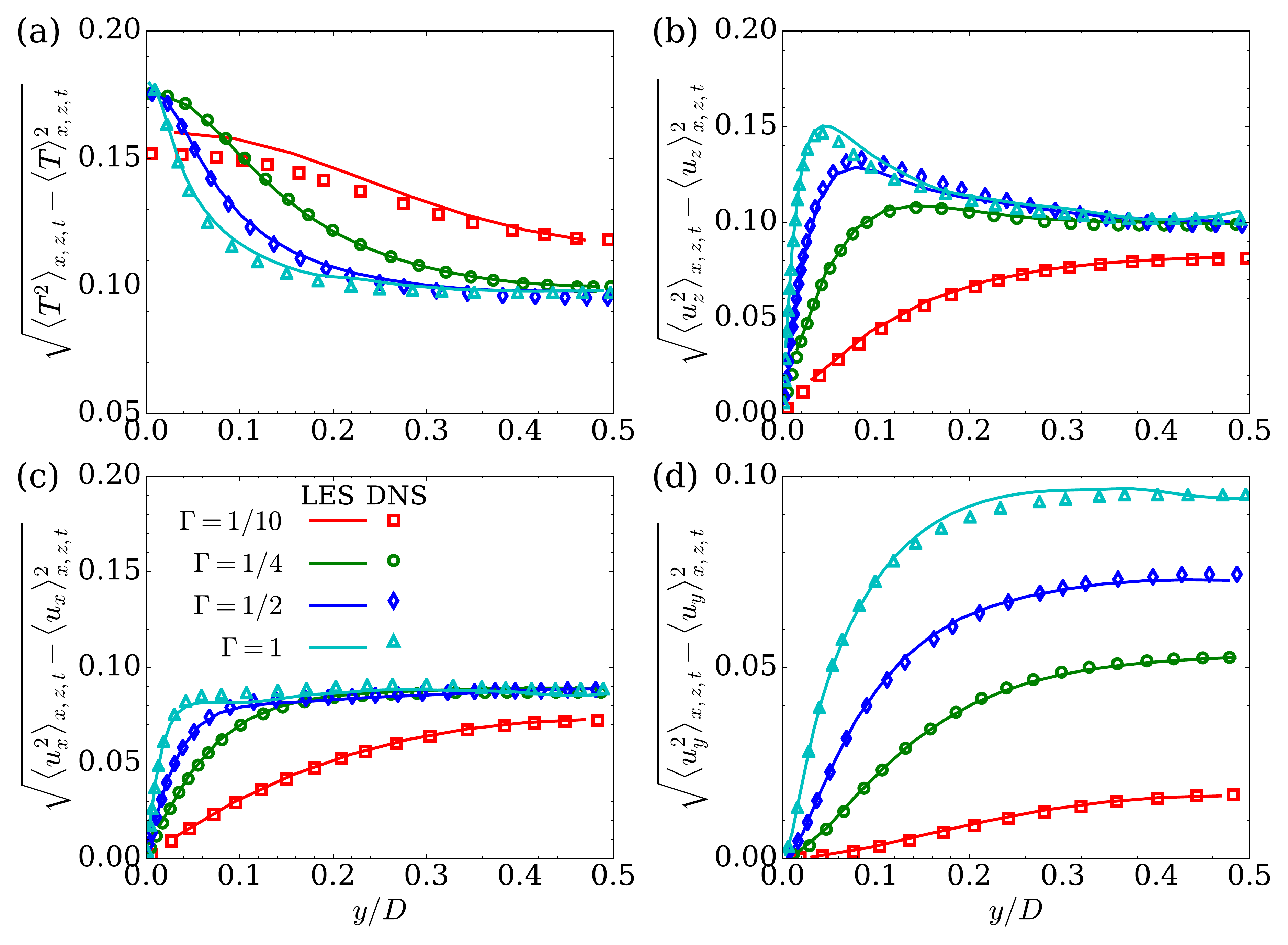}
    \caption{\label{fig:wagner_stats}
        Fluctuation profiles of (a) temperature T and the three components of velocity, (b) \(u_z\), (c) \(u_x\) and (d) \(u_y\)
        along y-axis for aspect ratios \(\Gamma = 1/10\), \(1/4\), \(1/2\) and 1.
        The data from DNS results of \citet{Wagner:PF2013} are also plotted as symbols corresponding to each aspect ratio.
        For the case with \(\Gamma = 1/10\), the LES uses just 16 points along the \(y\) direction, which leads to a slight mismatch in the temperature fluctuation profile close to the wall.
        All LES were performed at \(Ra = 10^7\) and \(Pr = 0.786\), as listed in Table~\ref{table:wag_hf_cases}.
            }
\end{figure}

We now conduct a set of LES experiments to verify that the SSV model can reproduce the fluctuation profiles of velocity and temperature correctly.
We use the results from the statistical analysis performed by \citet{Wagner:PF2013} to verify the LES predictions.
For the experiment, the \(Ra\) and \(Pr\) were fixed at \(10^7\) and 0.786 respectively.
Four values of \(\Gamma\) were chosen as listed in Table~\ref{table:wag_hf_cases}.
The quantities were averaged over both time \(t\) and in the vertical \(xz\)-plane (see Fig.~\ref{fig:domains}) to obtain the profiles as a function of the span-wise coordinate \(y\)
\be
    f(y) = \sqrt{\langle \psi^2(x, y, z, t) \rangle_{x, z, t} - \langle \psi(x, y, z, t) \rangle^2_{x, z, t}},
\ee
where \(\psi \in \{u_x, u_y, u_z, T\}\).
The above quantity is a measure of the standard deviation of the variable \(\psi\), and is related to the planar averaged fluctuation
\(\psi' = \psi - \langle \psi \rangle_{x, z, t}\)
as \(f(y) = \sqrt{\langle \psi' \psi' \rangle_{x, z, t}}\).
\be
    f(y) = \sqrt{\langle \psi' \psi' \rangle_{x, z, t}} = \sqrt{\langle (\psi - \langle \psi \rangle_{x, z, t})(\psi - \langle \psi \rangle_{x, z, t}) \rangle_{x, z, t}} = \sqrt{\langle \psi^2 \rangle_{x, z, t} - \langle \psi \rangle^2_{x, z, t}}. \nonumber
\ee

We compare the horizontal fluctuation profiles of LES (solid lines) against DNS (symbols) in Fig.~\ref{fig:wagner_stats}.
The good match between LES and DNS  in the aspect-ratio dependency of the statistics is a testimony that the subgrid-scale model captures the small scale physics well.
In Fig.~\ref{fig:wagner_stats}(c), we see that the LES correctly captures the local maxima close to the wall
in the fluctuation profiles of \(u_z\) for \(\Gamma \geq 1/4\).
\citet{Wagner:PF2013} attributes the lack of a similar maximum for \(\Gamma = 1/10\) to the merging of the two boundary layers at opposite walls along the \(y\)-direction.
There is a noticeable deviation in the temperature profile close to the wall for the case with \(\Gamma = 1/10\) in Fig.~\ref{fig:wagner_stats}(a),
but this may be attributed to the somewhat coarse grid resolution of approximately 16 points in the \(y\)-direction chosen for this case.
Despite such a low resolution, the SSV model reliably obtains the profiles of fluctuation for all the components of the velocity field for the same case.
The verification of the LES against DNS for these RBC flows lends confidence in the robustness of SSV model for thermal convection.
We next undertake LES of RBC at very high \(Ra\) and compare the Nusselt number and Reynolds number scaling laws.


\section{LES of RBC at high Rayleigh number}

Presently, we simulate Rayleigh B\'{e}nard Convection in a tall thin column.
In this case, the aspect ratio of the domain is defined as \(\Gamma = D/H = W/H\), where \(D\), \(W\) and \(H\) retain the same definitions as used before in Fig.~\ref{fig:domains}.
We fix \(\Gamma = 0.1\) for all the cases presented here.
We first perform the LES with adiabatic, no-slip walls confining the convective flow.
Subsequently, we investigate the effect of confinement on the structure and properties of the flow by replacing the side-walls with periodic boundary conditions.


\subsection{Adiabatic side-walls}
\label{sec:les_rbc}

With non-periodic boundaries, this case is similar to the DNS performed by \citet{Iyer:PNAS2020},
however we use a cuboidal domain instead of the cylindrical one used in the DNS.
We expect this modified geometry to have a marginal impact on the results and comparisons that we present next.
The full list of cases, with grid size, computed Reynolds and Nusselt numbers, and extent of simulation time in free-fall units are given in Table~\ref{table:hrbc_nper}.
In order to collect additional data against which the LES implementation can be verified,
we also performed a DNS at \(Ra = 10^{11}\) and \(Pr = 1.0\).
This is listed as an additional case B1 in the table.

\begin{table}
    \caption{\label{table:hrbc_nper}
        List of cases described in Section~\ref{sec:les_rbc} with details of Rayleigh number \(Ra\), grid points \(\mathrm{N}_x \times \mathrm{N}_y \times \mathrm{N}_z\), Nusselt number \(Nu\),
        Reynolds number \(Re\), and total simulation times in free-fall units, \(t_\mathrm{max}\).
    }
    \begin{tabular}{lccccr}
    Case        & \(Ra\)      & Grid                          & \(Nu\)            & \(Re\)                  & \(t_\mathrm{max}\)   \\
    \hline
    A1 (LES)    & \(10^{9}\)  & $32 \times 32 \times 256$     & \(54 \pm 6\)      & \(1880 \pm 202\)      & 320     \\
    A2 (LES)    & \(10^{10}\) & $32 \times 32 \times 256$     & \(105 \pm 6\)     & \(5826 \pm 597\)      & 340     \\
    A3 (LES)    & \(10^{11}\) & $64 \times 64 \times 512$     & \(212 \pm 11\)    & \(16580 \pm 1910\)    & 170     \\
    A4 (LES)    & \(10^{12}\) & $128 \times 128 \times 1024$  & \(497 \pm 23\)    & \(55860 \pm 7100\)    & 100     \\
    A5 (LES)    & \(10^{13}\) & $128 \times 128 \times 1024$  & \(1017 \pm 71\)   & \(196100 \pm 28900\)  & 100     \\
    A6 (LES)    & \(10^{14}\) & $512 \times 512 \times 4096$  & \(2304 \pm 33\)   & \(548000 \pm 4200\)   & 5       \\
    A7 (LES)    & \(10^{15}\) & $512 \times 512 \times 4096$  & \(3800 \pm 31\)   & \(1724200 \pm 8000\)  & 5       \\
    B1 (DNS)    & \(10^{11}\) & $256 \times 256 \times 2048$  & \(231 \pm 7\)     & \(17180 \pm 1040\)    & 30      \\
    \hline
    \end{tabular}
\end{table}

Compared to the DNS performed by \citet{Iyer:PNAS2020} the SSV model uses grids that are more than two orders of magnitude (\(\approx\) 120 times) smaller.
To meet the demands of wall-resolved LES, the grids are comparatively finer for \(Ra = 10^{14}\) and \(10^{15}\).
However, even in this case, our grids are 20 times smaller.
Moreover, for \(Ra\) up to \(10^{13}\), the LES has also been run for more units of free-fall time to obtain reliable time-averaged values.
In Appendix~\ref{app:time_series}, we briefly comment on this significant advantage of LES by examining the temporal variations of Nusselt and Reynolds numbers.
Experiments are usually conducted for a duration of hundreds of non-dimensional time-units to achieve statistical stationarity.
Meeting this requirement under the heavy computational burden of DNS is a major challenge.
Except for the two cases A6 and A7 which used the largest grids to resolve the extremely thin boundary layers, all our LES results are averaged for 100 units of time or more.
This demonstrates the computational advantage of LES that lends comparatively greater confidence in the stationarity of computed quantities.

\begin{figure}
    \centering
    \includegraphics[width=1.0\textwidth]{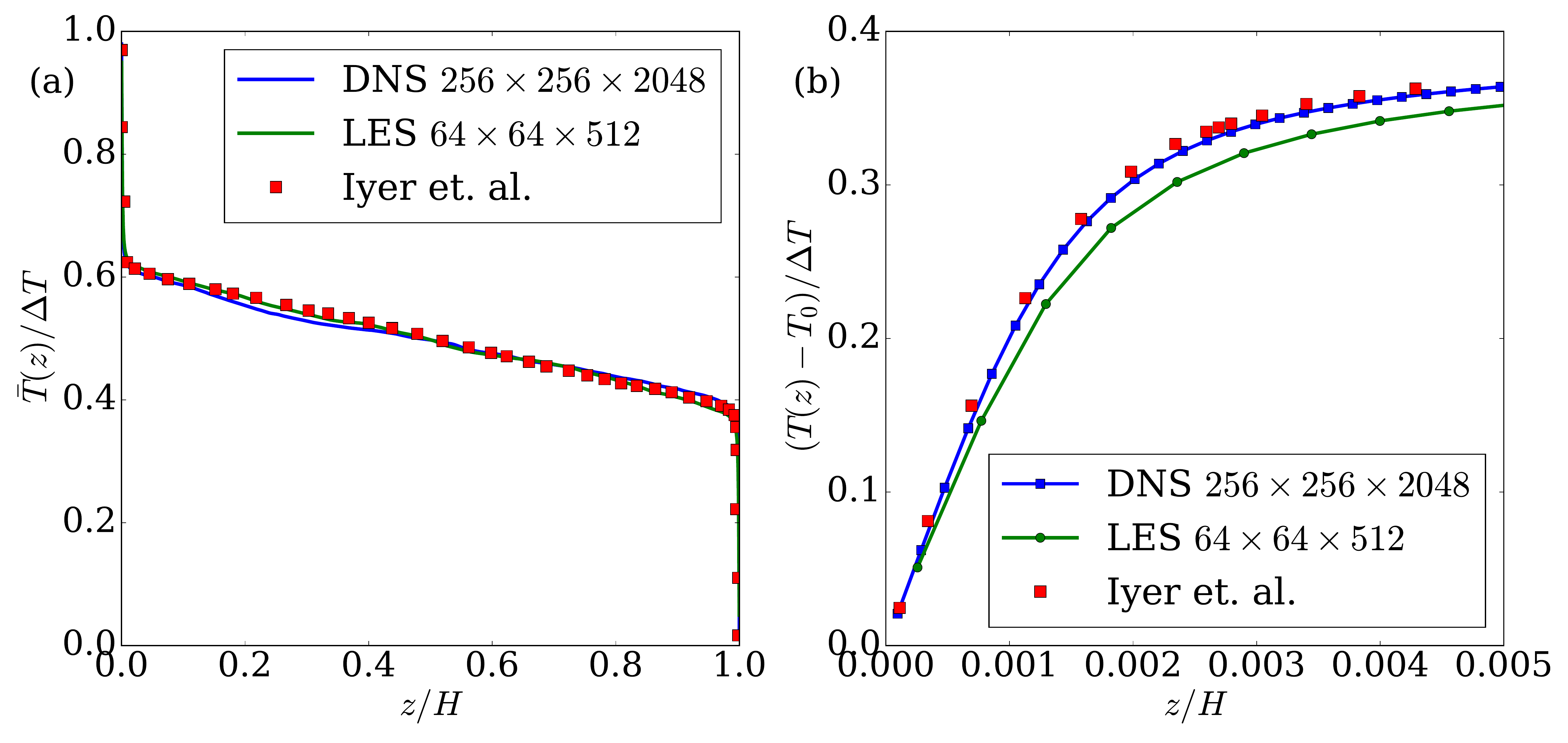}
    \caption{\label{fig:temp_profiles}
        Planar and time averaged profiles of dimensionless temperature against non-dimensionalized height for cases A3 and B1.
        We compare the profiles at a fixed \(Ra\) of \(10^{11}\) and \(Pr = 1\), taken from \citet{Iyer:PNAS2020}.
        Profiles are shown for
        (a) the full height of the cell from 0 to 1, and
        (b) within the boundary layer near the top and bottom wall-regions (combined through averaging).
            }
\end{figure}

Fig.~\ref{fig:temp_profiles} shows the dimensionless temperature (\(\bar{T}(z)/\Delta T\)) profiles plotted against the normalized cell-height (\(z/H\))
for cases A3 and B1.
The temperature field is averaged over the square cross-section and time for each \(z\).
Fig.~\ref{fig:temp_profiles}(a) indicates that the LES matches both the DNS results very well throughout the full range of the cell-height.
Within the thermal boundary close to the wall shown in Fig.~\ref{fig:temp_profiles}(b),
the temperature profile from LES follows the expected behaviour reasonably well even though there are 2-3 times fewer grid points
in the shear layer as compared to either of the two DNS results.

\begin{figure}
    \centering
    \includegraphics[width=1.0\textwidth]{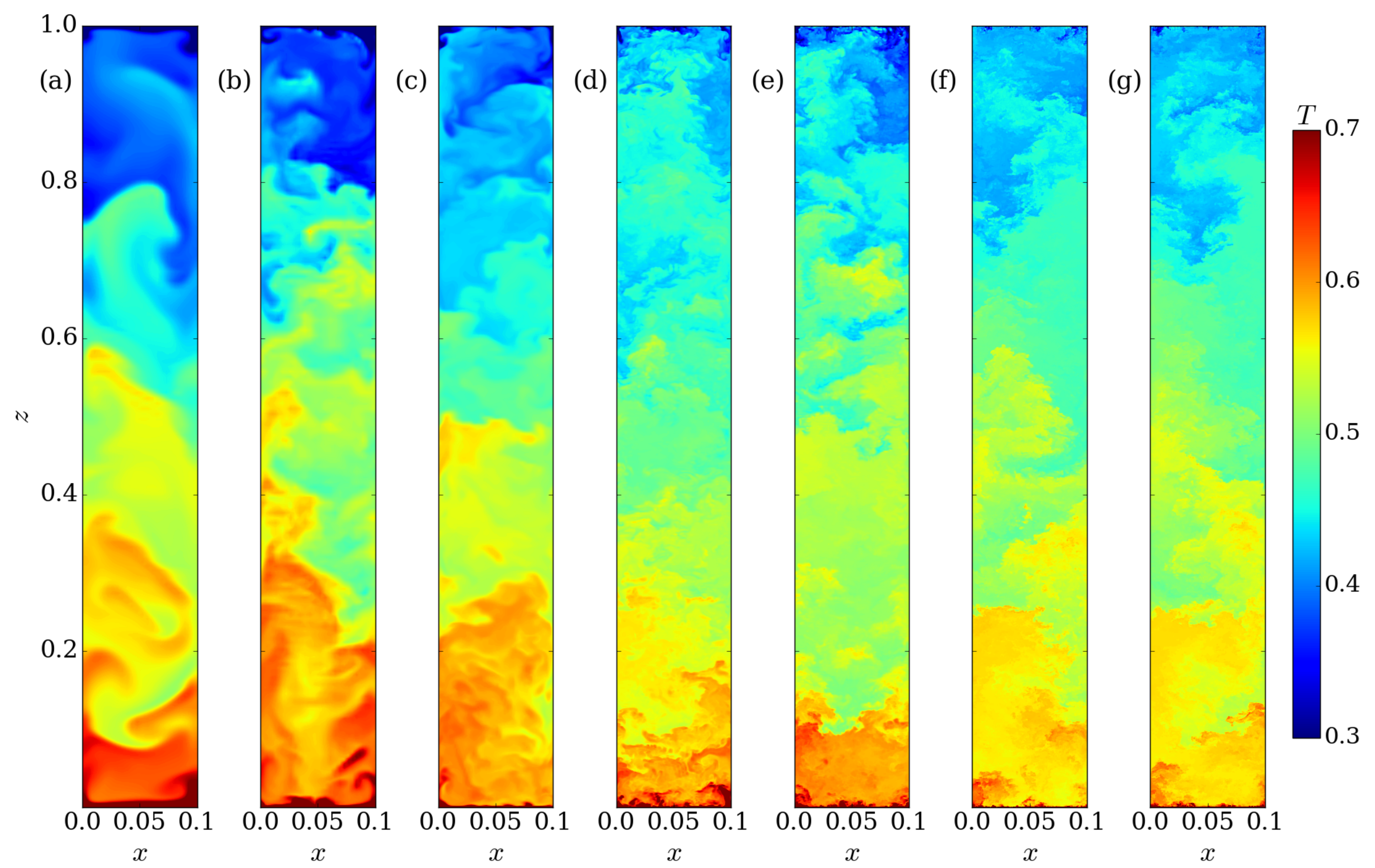}
    \caption{\label{fig:thin_col_temp}
        Contour plots of instantaneous temperature field on the mid \(y\)-plane at \(Pr = 1\) and (a) \(Ra = 10^9\),
        (b) \(10^{10}\), (c) \(10^{11}\), (d) \(10^{12}\), (e) \(10^{13}\), (f) \(10^{14}\), and (g) \(10^{15}\).
        The decreasing thickness of the thermal boundary layers with increasing \(Ra\) is noticeable.
        Predictably, the flow structures also grow increasingly fine and the bulk of fluid in the center at mean temperature grows wider as well.
            }
\end{figure}

\begin{figure}
    \centering
    \includegraphics[width=1.0\textwidth]{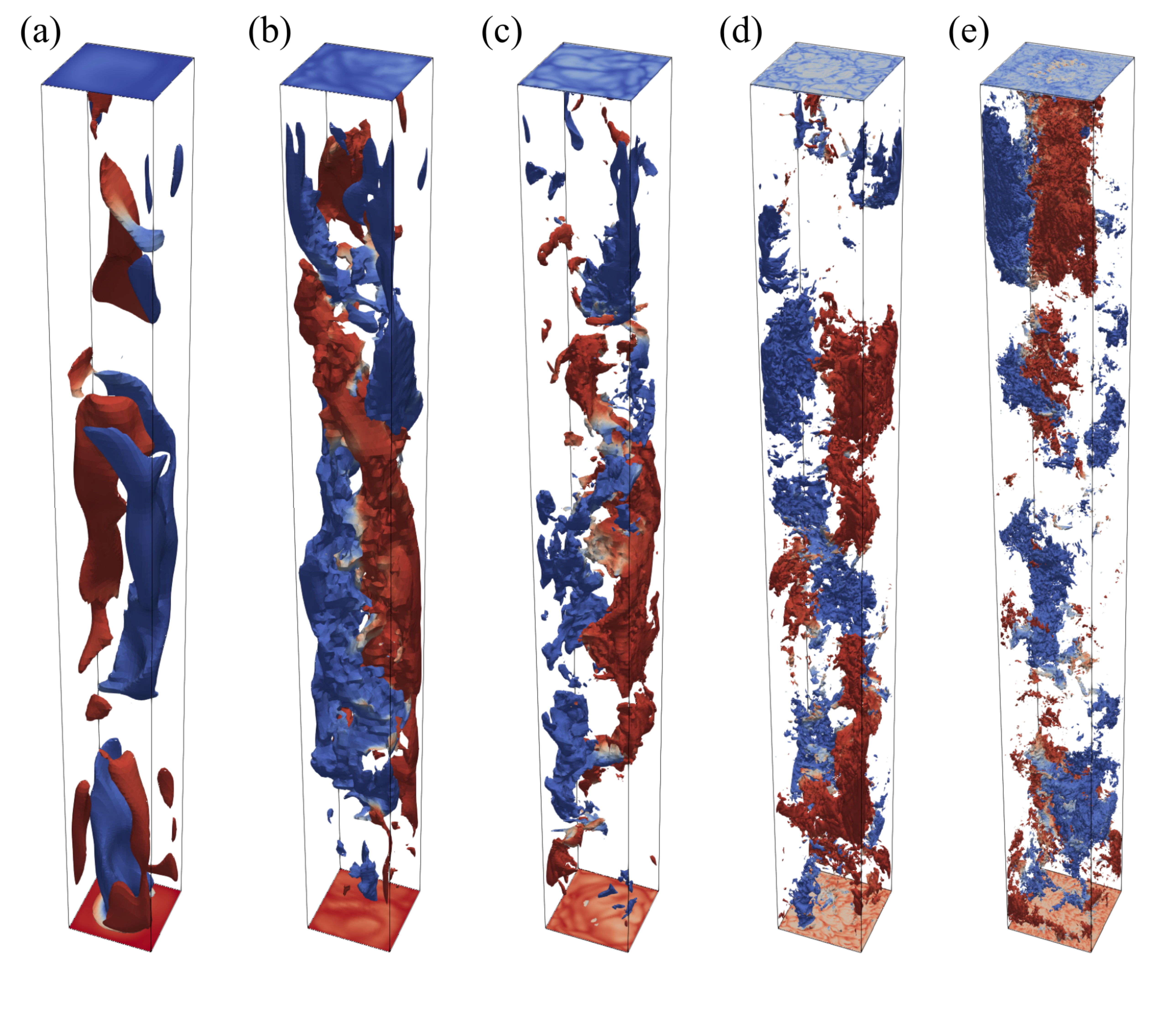}
    \caption{\label{fig:thin_col_vz_nper}
        Iso-volumes of velocity magnitude, \(|\mathbf{u}|\), coloured by the \(z\)-component of velocity, \(u_z\) at \(Pr = 1\) and (a) \(Ra = 10^9\),
        (b) \(10^{10}\), (c) \(10^{11}\), (d) \(10^{12}\), and (e) \(10^{13}\).
        The red-coloured volumes indicate packets of fluid rising up, whereas the blue-coloured volumes are patches of fluid falling down from the cold plate.
        The two iso-surfaces of hot and cold fluids are entwined in a double helix.
            }
\end{figure}

Fig.~\ref{fig:thin_col_temp} shows the vertical cross-sections of the temperature field on the \(xz\)-plane situated at \(y = 0.05\).
Both the decreasing thickness of boundary layers and the increasing complexity of the flow is apparent in the frames from left to right
(arranged according to increasing Rayleigh number from \(10^9\) to \(10^{15}\)).
The subgrid model also captures another interesting aspect of the large-scale flow organization.
In \citet{Iyer:PNAS2020}, it was noted that the convective flow within a thin cylindrical column forms a helical structure similar to a barber pole.
For larger aspect ratios as seen in the previous section, the presence of large-scale circulation (LSC) in the flow is distinctly apparent.
However, due to the highly confined space in a thin column, the recirculating flow comprised of hot fluid rising from the bottom plate
and cold fluid falling from the top plate twists and arranges itself into a pair of entwined helices.
This is highlighted in Fig.~\ref{fig:thin_col_vz_nper}, which shows the iso-surfaces of velocity magnitude in the range of 0.07 to 0.13
for the cases A1 to A5.
The iso-surfaces are coloured by the \(z\)-component of velocity, \(v_z\), so that the red surfaces enclose fluid packets rising up the cell,
while the blue volumes are patches of cold fluid descending down the column.
Interestingly, although \citet{Iyer:PNAS2020} observed this behaviour in a column with a circular cross-section,
our LES also reproduces this structure in a column with a square cross-section.
This indicates that the helical structure of the macroscopic flow field is a universal feature enforced by the aspect ratio of the domain,
and not strictly influenced by the particular choice of the cross-section.
However an additional facet in the formation of such a flow is the confinement produced by the adiabatic no-slip walls on the sides.
In Section~\ref{sec:rbc_per}, we investigate the effect of removing this restriction on the characteristics of the flow-field.

The presence of vertically stacked rolls in thermal convection in boxes of \(\Gamma < 1\) has been observed previously \cite{vanderPoel:PRE2011}
in numerical experiments of RBC in both two and three dimensions \cite{vanderPoel:PRE2014}.
From Fig.~\ref{fig:thin_col_vz_nper}, we deduce the origin of vertically stacked rolls from the helical large-scale flow.
Each roll is formed at the intersection of a rising plume of hot fluid, and a falling packet of cold fluid,
such that the components of velocity from the double helices, when projected onto the vertical mid-plane, form the circulating flow.
However, we surmise that the stacked rolls observed in 2D simulations of RBC must arise through an alternative route.

\begin{figure}
    \centering
    \includegraphics[width=1.0\textwidth]{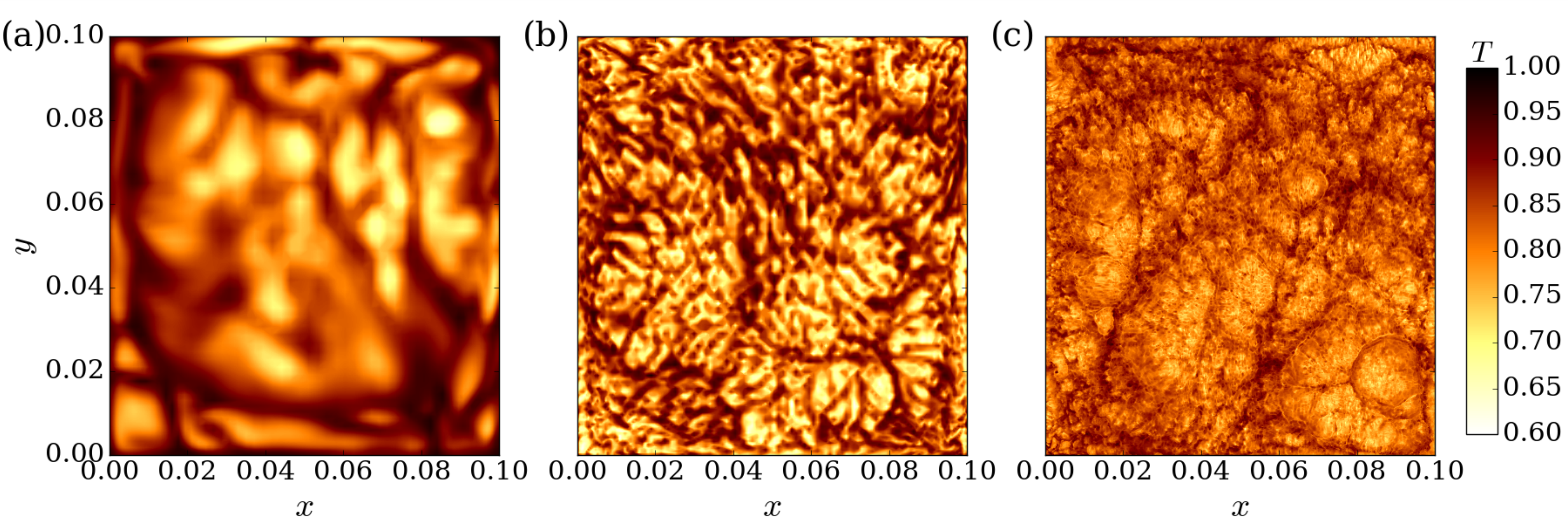}
    \caption{\label{fig:thin_col_bot_temp}
        Contour plots of the temperature field on a horizontal cross-section inside the boundary layer at the bottom plate.
        The \(Ra\) and height of the plane of cross-section are:
        (a) \(Ra = 10^{11}\) at \(z = 0.0008 H\),
        (b) \(Ra = 10^{13}\) at \(z = 0.0004 H\) and
        (c) \(Ra = 10^{15}\) at \(z = 0.00005 H\).
            }
\end{figure}

The existence of sheet plumes in RBC has been observed in numerous experimental studies of RBC \cite{Baburaj:JFM2005, Funfschilling:PRL2004}.
\citet{Shishkina:JFM2008} performed a detailed numerical investigation of these coherent structures through both DNS and LES.
In Fig.~\ref{fig:thin_col_bot_temp} we plot the temperature field on a horizontal (\(xy\)) plane, situated within the boundary layer near the bottom plate of the column.
The three snapshots, taken at Rayleigh numbers (a) \(10^{11}\), (b) \(10^{13}\), and (c)  \(10^{15}\), show the structure of convective flow near the walls.
Due to the decreasing thickness of the boundary layer with increasing \(Ra\), the heights of the planes also decrease from left to right of the figure.
We see that the granularity of the flow increases with increase in the Rayleigh number,
as manifested by the finer structure of the plumes rising from the bottom plate (marked by the dark regions in the plot).

\begin{figure}
    \centering
    \includegraphics[width=0.48\textwidth]{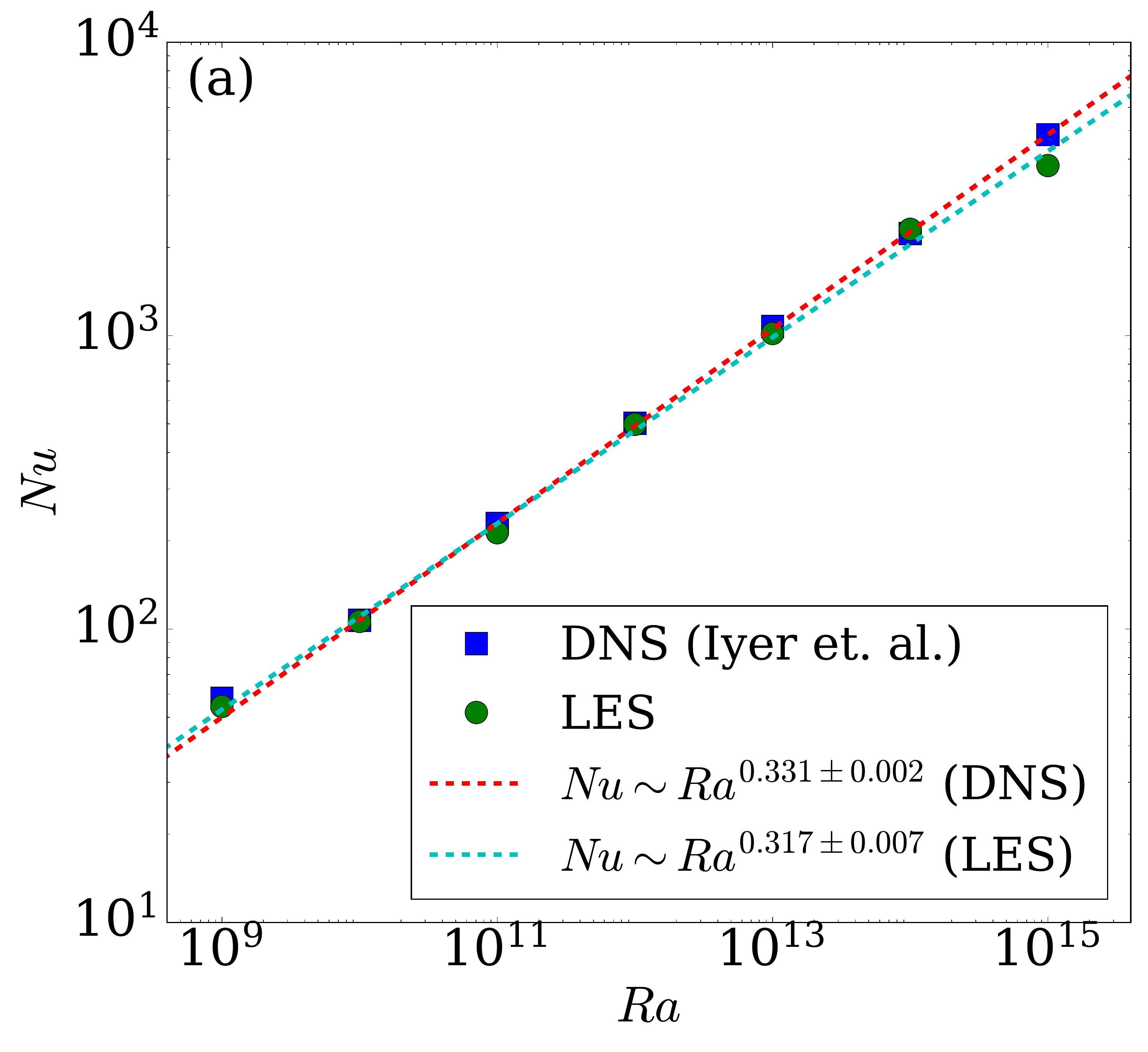}
    \includegraphics[width=0.48\textwidth]{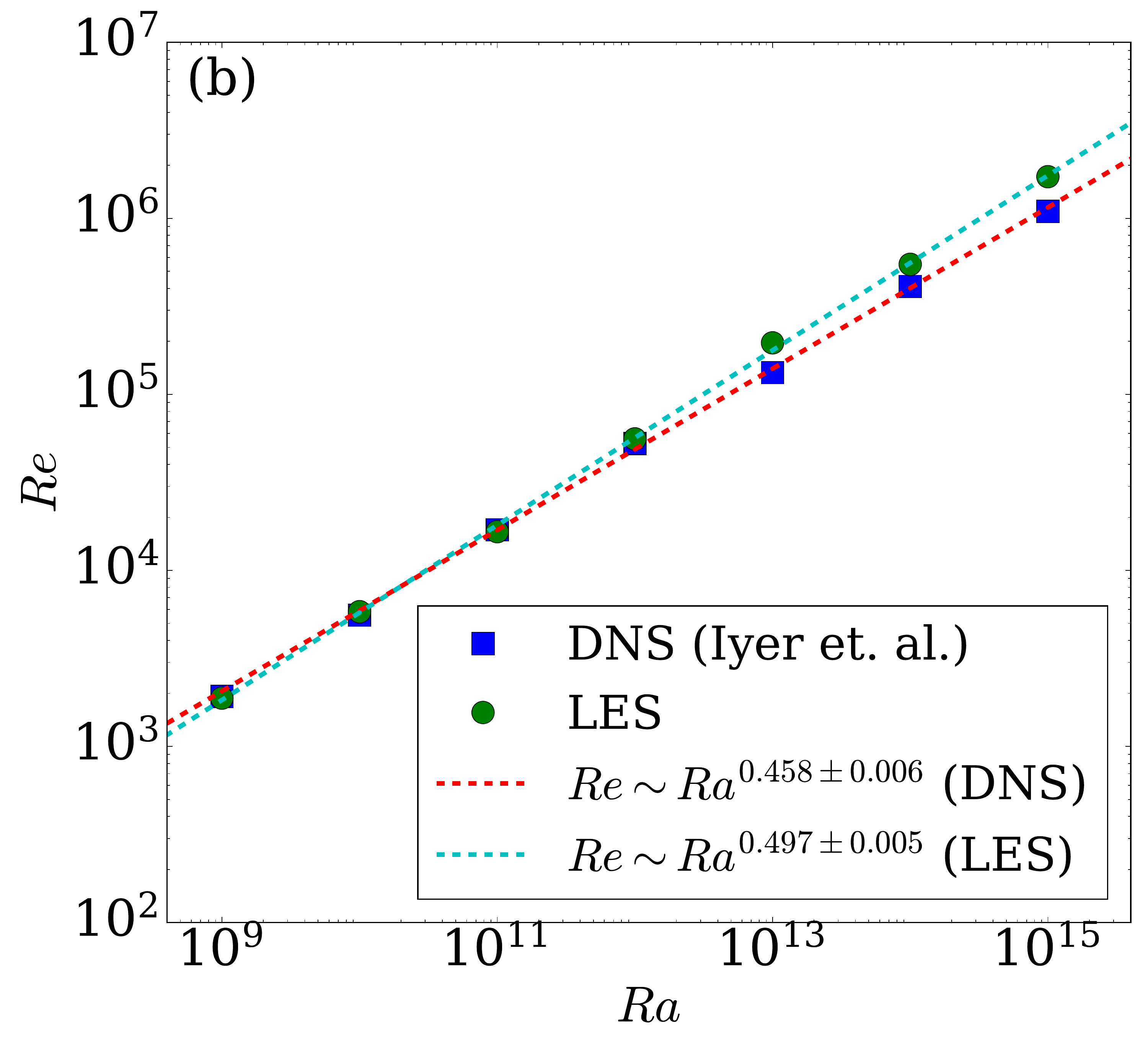}
    \caption{\label{fig:re_nu_scale}
        The scaling of global quantities \(Nu\) (left) and \(Re\) (right) with \(Ra\).
        The LES data is represented by green circles and the DNS data from \citet{Iyer:PNAS2020} are plotted with blue squares.
        The corresponding scaling laws derived using least-squares fit are plotted as dashed cyan and red lines, respectively.
        (a) The LES data agree very closely with the classical scaling of \(Nu\), and the power law fit gives \(Nu = (0.074 \pm 0.009) \times Ra^{0.317 \pm 0.007}\).
        (b) correspondingly, \(Re\) in our LES scales as \(Re = (0.062 \pm 0.006) \times Ra^{0.497 \pm 0.005}\).
            }
\end{figure}

We conclude this section on LES results with an analysis of the scaling laws predicted by the stretched-vortex model.
In Figs.~\ref{fig:re_nu_scale}(a) and \ref{fig:re_nu_scale}(b), we plot the Nusselt and Reynolds numbers respectively against Rayleigh number ranging from \(10^9\) to \(10^{15}\).
We also show the corresponding values from the DNS performed by \citet{Iyer:PNAS2020} for comparison.
For both \(Nu\) and \(Re\), we use a least-squares fit to obtain their respective power laws, which are plotted with dashed lines in the two frames.
We see that according to our LES, Nusselt number scales as \(Nu = (0.074 \pm 0.009) \times Ra^{0.317 \pm 0.007}\).
Compared to the scaling exponent of \(0.331 \pm 0.002\) obtained by \citet{Iyer:PNAS2020} for a similar configuration,
the value computed from our LES is in agreement with a 4\% margin.

Similarly, we obtain the Reynolds number scaling as \(Re = (0.062 \pm 0.006) \times Ra^{0.497 \pm 0.005}\) from the LES data.
Interestingly, the scaling exponent for \(Re\) is very close to the \(1/2\) scaling which has been reported previously in literature.
For instance, \citet{Scheel:PRF2017} performed direct simulations of RBC in a cylinder of aspect-ratio 1, and obtained a scaling of \(Re \sim Ra^{0.49 \pm 0.01}\) for a fluid with \(Pr = 0.7\).
However, an analysis of large-scale quantities performed by \citet{Pandey:PF2016} have shown that
Reynolds number will scale as \(Re \sim Ra^{0.38 \pm 0.01}\) for certain regimes of turbulence in RBC \cite{Verma:book:BDF}.
We will revisit the scaling of these large-scale quantities in the next section.
Also, our scaling exponent for \(Re\) differs by a somewhat greater margin of 8.5\% from the \(Re \sim Ra^{0.458 \pm 0.006}\) obtained by \citet{Iyer:PNAS2020}.
This difference is mainly due to the higher values of \(Re\) in the LES at high Rayleigh numbers.
It is possible that the difference in geometry is a reason for this difference.
A circular cross-section of diameter 0.1 units tends to have a greater confining effect than a square cross-section of width 0.1 units.
This can lead to a slight reduction in overall kinetic energy of the flow.


\subsection{Periodic side-walls}
\label{sec:rbc_per}

\begin{figure}
    \centering
    \includegraphics[width=1.0\textwidth]{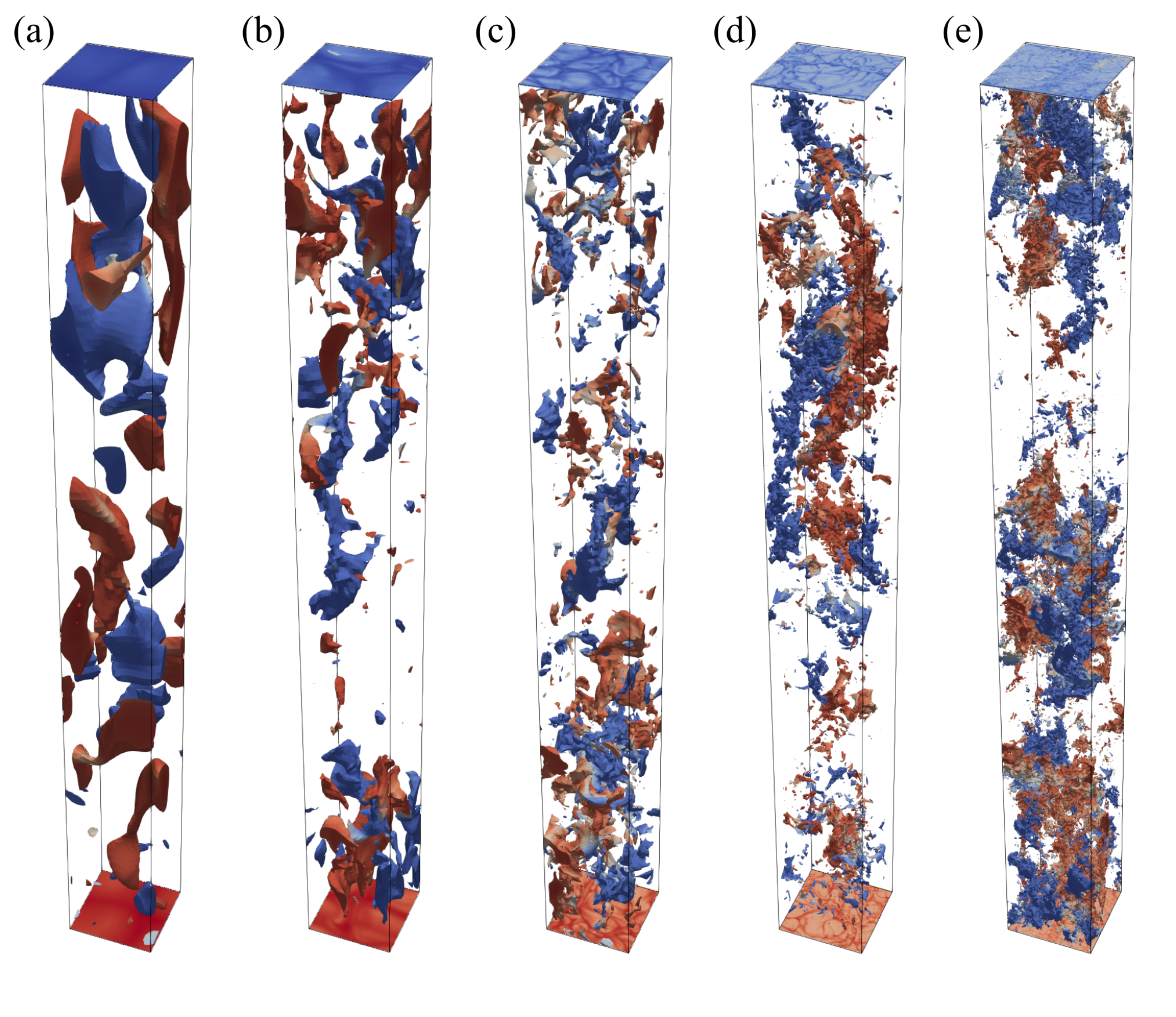}
    \caption{\label{fig:thin_col_vz_per}
        Iso-volumes of velocity magnitude, \(|\mathbf{u}|\), coloured by the \(z\)-component of velocity, \(u_z\) at \(Pr = 1\) and (a) \(Ra = 10^9\),
        (b) \(10^{10}\), (c) \(10^{11}\), (d) \(10^{12}\), and (e) \(10^{13}\).
        The red-coloured volumes indicate packets of fluid rising up, whereas the blue-coloured volumes are patches of fluid falling down from the cold plate.
        Unlike the flow seen in Fig.~\ref{fig:thin_col_vz_nper}, here the helical structure has mostly collapsed and the mixing of hot and cold fluid is more chaotic.
            }
\end{figure}

\begin{table}
    \caption{\label{table:hrbc_per}
        List of cases described in Section~\ref{sec:rbc_per} with details of Rayleigh number \(Ra\), grid points \(\mathrm{N}_x \times \mathrm{N}_y \times \mathrm{N}_z\), Nusselt number \(Nu\),
        Reynolds number \(Re\), and total simulation times in free-fall units, \(t_\mathrm{max}\).
    }
    \begin{tabular}{lccccr}
    Case        & \(Ra\)      & Grid                          & \(Nu\)            & \(Re\)                  & \(t_\mathrm{max}\) \\
    \hline
    C1 (LES)    & \(10^{9}\)  & $32 \times 32 \times 256$     & \(36 \pm 3\)      & \(1615 \pm 128\)      & 200   \\
    C2 (LES)    & \(10^{10}\) & $32 \times 32 \times 256$     & \(75 \pm 7\)      & \(4624 \pm 372\)      & 200   \\
    C3 (LES)    & \(10^{11}\) & $64 \times 64 \times 512$     & \(179 \pm 17\)    & \(13150 \pm 1010\)    & 125   \\
    C4 (LES)    & \(10^{12}\) & $128 \times 128 \times 1024$  & \(392 \pm 24\)    & \(40780 \pm 1890\)    & 100   \\
    C5 (LES)    & \(10^{13}\) & $128 \times 128 \times 1024$  & \(814 \pm 45\)    & \(132500 \pm 8000\)   & 100   \\
    \hline
    \end{tabular}
\end{table}

We now simulate Rayleigh B\'{e}nard Convection in a tall thin column with the same aspect ratio of \(1/10\) as used in the previous section,
but with the adiabatic, no-slip sidewalls replaced with periodic boundary conditions.
Earlier, the confinement of the flow imposed by the walls had resulted in a specific organization of the large-scale flow
within the column.
By eliminating this constraint, we seek to understand the change in structure of the flow and its effects on
momentum and heat transport between the two plates.
We performed this set of simulations for a slightly reduced range of \(Ra\) from \(10^9\) to \(10^{13}\) at a fixed \(Pr = 1\) as described in the previous section.
Table~\ref{table:hrbc_per} lists the set of LES described in this section.

\begin{figure}
    \centering
    \includegraphics[width=0.7\textwidth]{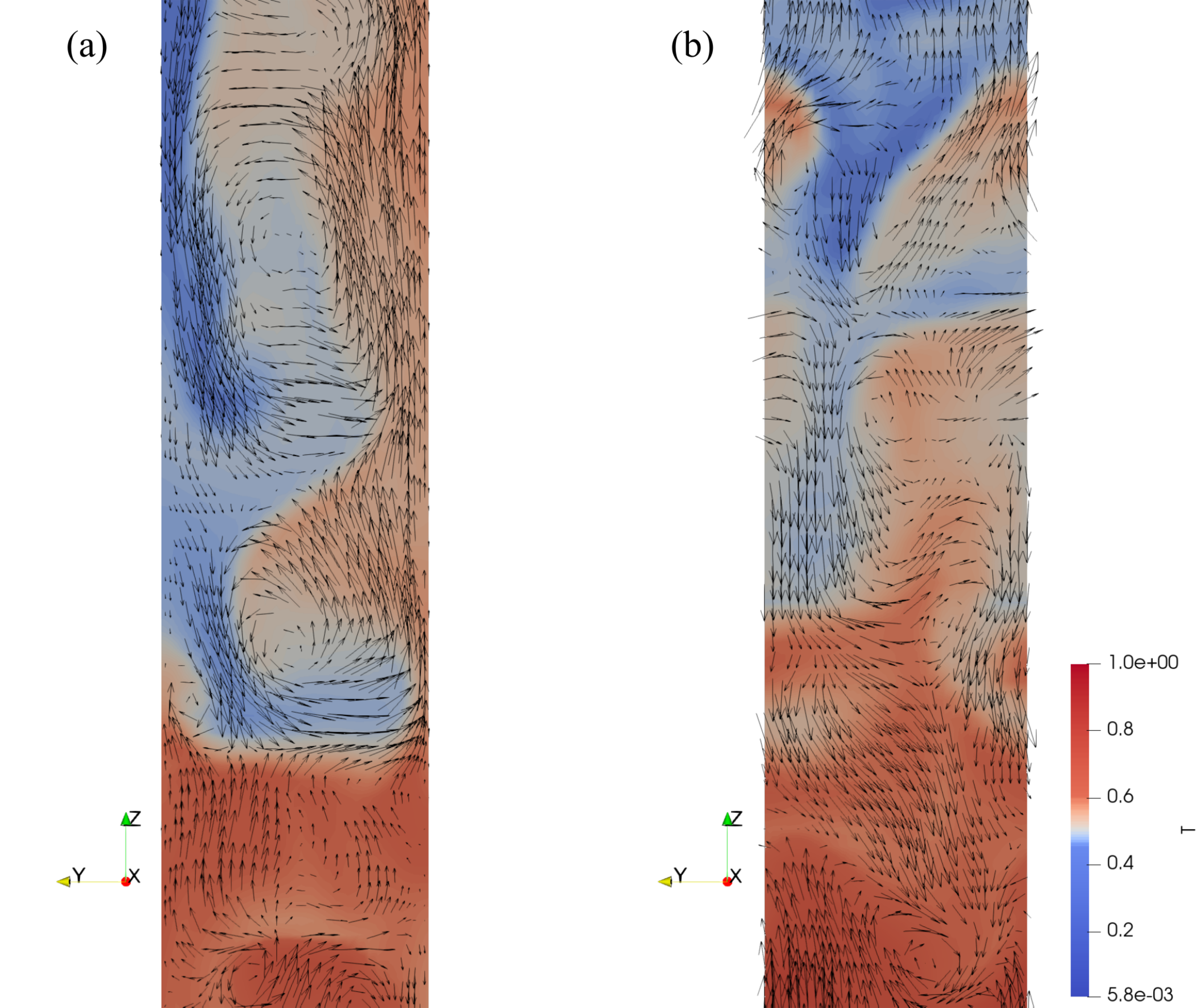}
    \caption{\label{fig:stacked_rolls}
        Contours of instantaneous temperature field, \(T\), along with \((u_z, u_y)\) velocity vectors on the mid \(yz\)-plane at \(Ra = 10^9\) and \(Pr = 1\).
        On the left we show the flow-field in the non-periodic case, whereas the right frame corresponds to the periodic side-wall simulation.
        We observe that when the convective flow is not confined, there is no evidence of vertically stacked rolls,
        leading to decreased transport of heat and momentum.
            }
\end{figure}

Earlier, when the convecting flow in the slender column was confined by side-walls,
the large-scale structure of the flow had reorganized itself into a pair of entwined double-helix (see Fig.~\ref{fig:thin_col_vz_nper}).
However, with periodic walls in the horizontal direction, the flow is no longer restricted, as seen in Fig.~\ref{fig:thin_col_vz_per}.
As a result, the rising hot fluid and falling cold fluid no longer forms a coherent structure.
In Fig.~\ref{fig:stacked_rolls}, we compare this effect at \(Ra = 10^9\) by plotting the instantaneous temperature and velocity field on
the mid \(yz\)-plane for both the non-periodic (left frame) and periodic (right frame) simulations.
With the no-slip walls confining the fluid, the flow forms stacked rolls as a consequence of the helical structure of the flow observed earlier in Fig.~\ref{fig:thin_col_vz_nper}.
However, the rolls are absent when the fluid is allowed to flow freely across the side-walls.
Consequently in this case, the hot and cold fluids mix chaotically, bereft of a coherent updraft and downdraft to transport heat and momentum.
This leads to a reduced efficiency in the transport of these quantities.
This can be a potential cause for the depressed values in \(Re\) and \(Nu\) observed when the side-walls are removed.

\begin{figure}
    \centering
    \includegraphics[width=0.48\textwidth]{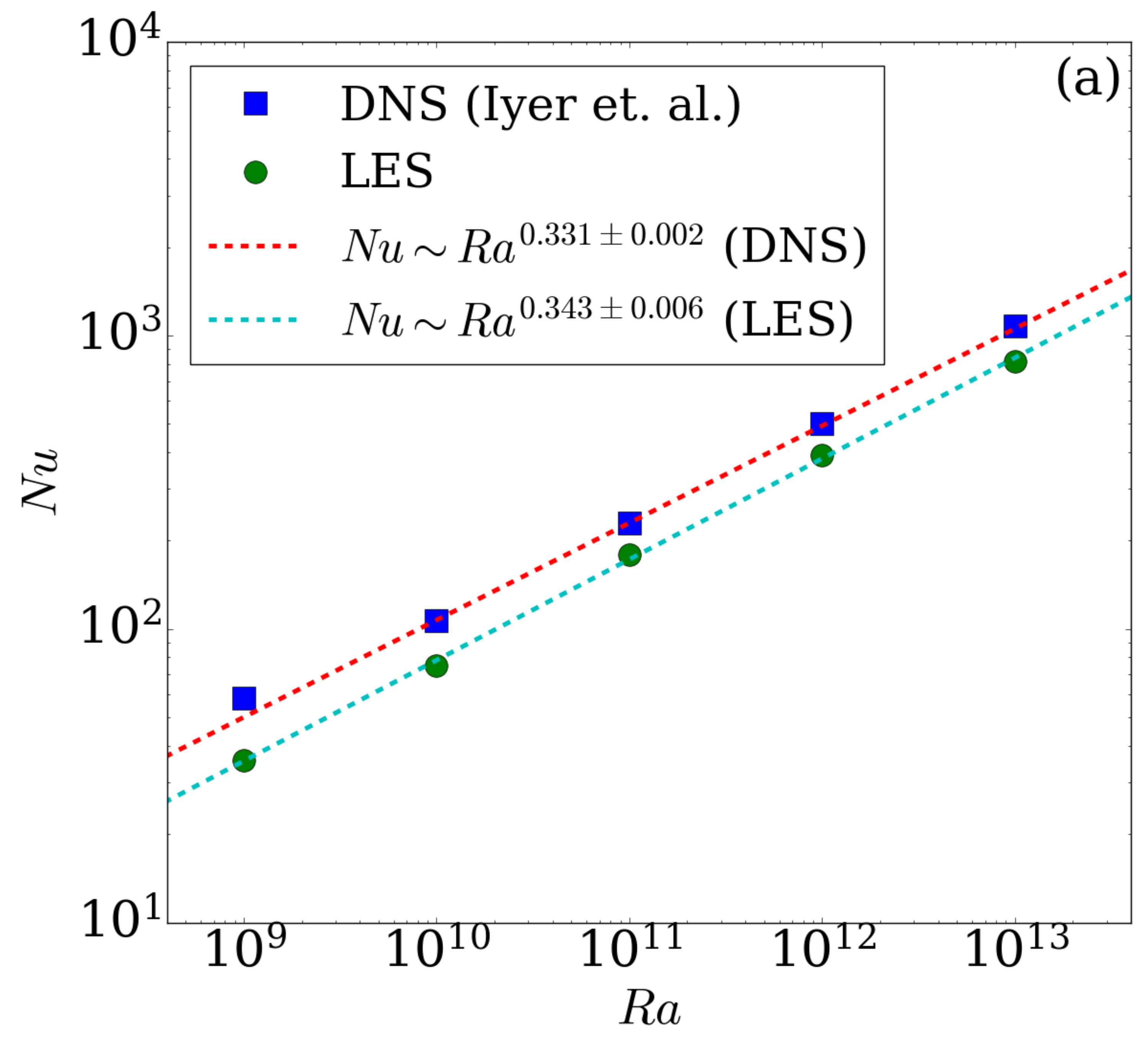}
    \includegraphics[width=0.48\textwidth]{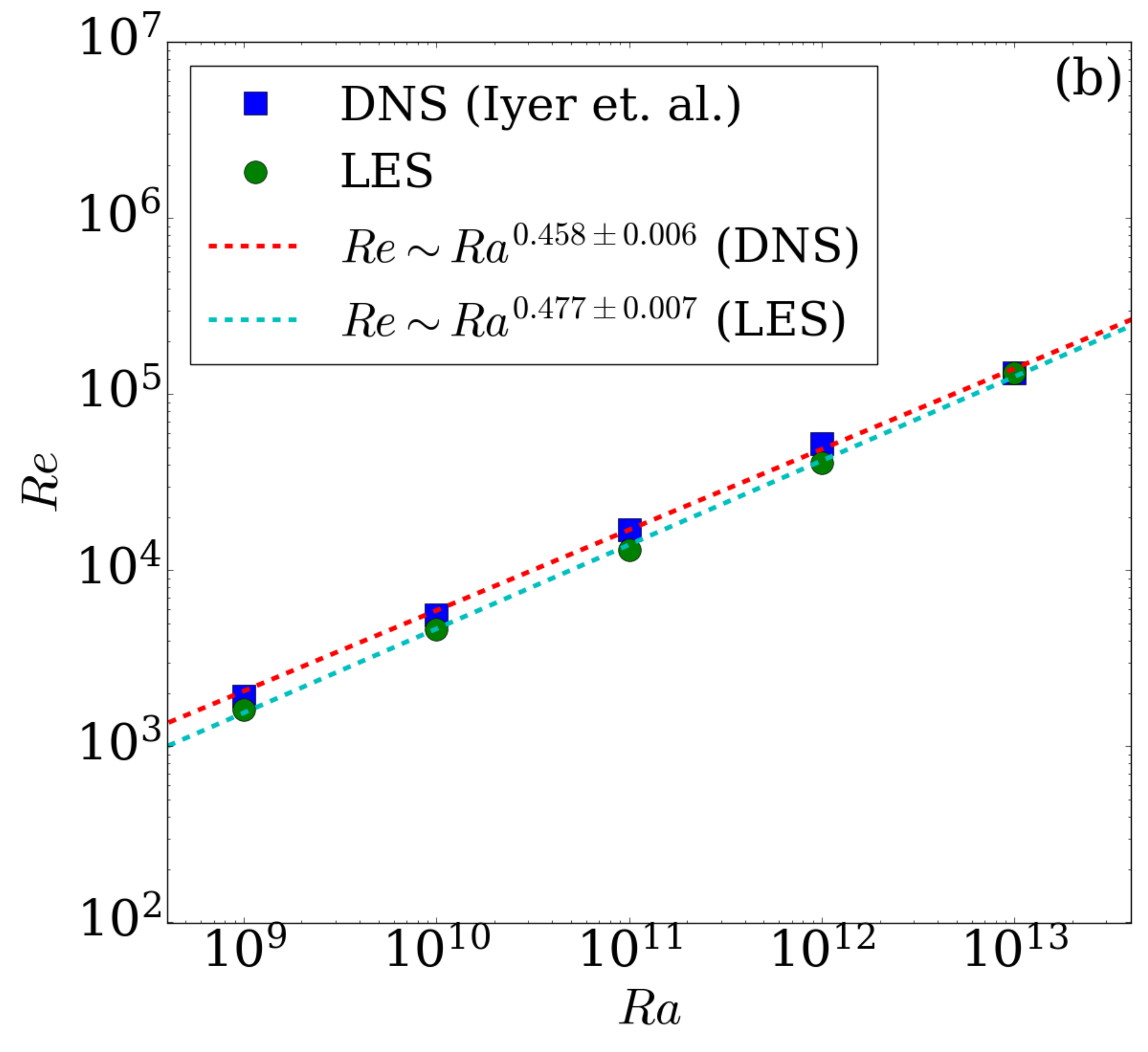}
    \caption{\label{fig:re_nu_scale_per}
        The scaling of global quantities \(Nu\) (left) and \(Re\) (right) with \(Ra\).
        The LES data is represented by green circles and the DNS data from \citet{Iyer:PNAS2020} are plotted with blue squares.
        The corresponding scaling laws derived using least-squares fit are plotted as dashed lines of cyan and red respectively.
        (a) The LES data agree very closely with the classical scaling of \(Nu\), and the power law fit gives \(Nu = (0.029 \pm 0.006) \times Ra^{0.343 \pm 0.006}\).
        (b) correspondingly, \(Re\) in our LES scales as \(Re = (0.078 \pm 0.008) \times Ra^{0.477 \pm 0.007}\).
            }
\end{figure}

Another interesting point to note is that the removal of side walls does not affect the scaling exponents of \(Re\) and \(Nu\) significantly.
Only the prefactors of the scaling laws are readjusted in the case of periodic boundaries.
In Fig.~\ref{fig:re_nu_scale_per}, we plot the scaling laws for \(Re\) and \(Nu\) in the absence of side-walls.
We obtain the power-law scalings of both quantities using least-squares fits as before.
The scaling laws for \(Nu\) for adiabatic walls and periodic walls are summarized as follows.
\begin{align}
    Nu &= (0.074 \pm 0.009) \times Ra^{0.317 \pm 0.007} &\text{(No-slip)}, \\
    Nu &= (0.029 \pm 0.006) \times Ra^{0.343 \pm 0.006} &\text{(Periodic)}.
\end{align}
Correspondingly, \(Re\) scales as
\begin{align}
    Re &= (0.062 \pm 0.006) \times Ra^{0.497 \pm 0.005} &\text{(No-slip)}, \\
    Re &= (0.078 \pm 0.008) \times Ra^{0.477 \pm 0.007} &\text{(Periodic)}.
\end{align}
We note that the scaling exponent for \(Nu\) is elevated by 8.2\% for the case with periodic side-walls.
However, the exponent for \(Re\) is decreased by 4\% as compared to the case with adiabatic walls.
\citet{Verma:book:BDF} notes that for a given value of Prandtl number, the convective flow in RBC makes a series of transitions from laminar regime to turbulent regime.
In the case of \(Pr = 1\) as presented here, the exponent \(\beta\) of \(Re \sim Ra^\beta\) will transition
from 0.6 (laminar regime), to 0.5 (intermediate regime), to 0.38 (turbulent regime I) and back to 0.5 (turbulent regime II).
The scaling laws from both the LES and DNS indicate that at high Rayleigh numbers (\(Ra > 10^{11}\)), the flow is closer to turbulent regime II.
However, \citet{Iyer:PNAS2020} notes that \(\beta = 0.5\) at high \(Ra\) ``does not necessarily herald the ultimate convection state'',
adding that a similar exponent has been observed in experiments in a unit aspect ratio cylinder \cite{Scheel:PRF2017}.
Similarly \citet{Bhattacharya:PF2021}, when extending the model by \citet{Grossmann:PRL2001} for predicting \(Nu\) and \(Re\),
observed that for an aspect-ratio of 1, \(Re\) scales as \(Re \sim Ra^{0.5}\) for moderate Prandtl numbers.

In the case of Nusselt number scaling, both periodic and non-periodic simulations have exponents close to the standard \(1/3\) scaling, similar to the results of \citet{Iyer:PNAS2020}.
Moreover, for the case with no-slip walls, the prefactor of 0.074 is very close to the theoretically predicted value of 0.073 by \citet{Malkus:PRSA1954} and \citet{Spiegel:ARAA1971}.
We now conclude the paper.


\section{Conclusions}
\label{sec:conclusions}

The present work realizes a novel type of large-eddy simulation model for numerical studies of Rayleigh-B\'{e}nard convection.
We implement the stretched spiral-vortex (SSV) model to calculate the subgrid-scalar flux of temperature in our LES.
We first compare the results from the LES against the DNS performed by \citet{Wagner:PF2013} at moderate to high \(Ra\) of \(10^6\) to \(10^9\).
The subgrid model was able to reproduce the effect of aspect ratio of the box on the structure and properties of convective flow.
This was further corroborated by modal analysis of the velocity field taken at the vertical mid-plane of the box.
In the final step of verification, we compared the fluctuation profiles of velocity and temperature in the box, and obtained a very good agreement with the DNS profiles.

Our primary goal in implementing an LES model for RBC is to probe the physics of convective flow at very high Rayleigh numbers, which is a field of active research.
To this end, we performed LES of convection at high \(Ra\) in a thin columnar box of aspect ratio 0.1, similar to the DNS experiments reported by \citet{Iyer:PNAS2020}.
Our simulations span 6 decades of Rayleigh number from \(10^9\) to \(10^{15}\).
We study the variation of Nusselt and Reynolds numbers with \(Ra\), and obtain a scaling law which is in good agreement with previous experimental results of RBC.

For convection in slender columns, the SSV model also reproduced the large-scale organization of the flow in the form of a pair of entwined helices.
This flow structure is more apparent at moderate \(Ra\) of \(10^9\), and it becomes less coherent as the \(Ra\) increases.
The velocity projected by this coherent structure on the vertical mid-plane of the cavity can be seen to form a set of vertically stacked rolls,
previously observed in simulations of RBC in boxes of low aspect ratio.

We also observe that the removal of the confining side-walls by using periodicity along the two horizontal directions leads to the disruption of the large-scale flow structure.
The resultant chaotic flow leads to a reduced efficiency of momentum and heat transport.
This is also indicated by the reduced values \(Nu\) and \(Re\) for the periodic case, although
the scaling exponents for both these parameters are not altered significantly.

An additional step towards attaining yet higher values of Rayleigh numbers to probe the existence of ultimate regime is to implement a thermal wall-model.
\citet{Chung:JFM2009} demonstrated that the SSV model can be adapted to compute the slip velocity at a lifted virtual wall which covers the no-slip boundaries.
This greatly alleviates the need for high grid resolution near no-slip walls.
This type of wall-modeled LES (WMLES) has proven to be very effective in numerous simulations of hydrodynamic turbulence \cite{Gao:JFM2019, Cheng:JFM2020, Gao:JFM2020}.
By demonstrating the efficacy of SSV model for RBC, we now aim to extend the wall model and approximate the thermal flux at a lifted virtual wall.
The resolution requirements for simulating RBC at \(Ra\) of the order \(10^{15}\) will be relaxed significantly by the introduction of such a thermal wall model.
This paper demonstrates the robustness of SSV model for LES of RBC, and thus presents a very promising tool in the search for ultimate regime in thermal convection.

\begin{acknowledgments}
Our numerical simulations were performed on Cray XC40 (Shaheen II) of KAUST supercomputing laboratory, Saudi Arabia, through Project k1416.
Some simulations were also performed on the High Performance Computing facility at IIT Kanpur, funded by the DST and IITK.
\end{acknowledgments}


\appendix


\section{Numerical Method}
\label{app:num_meth}

The low-storage third-order Runge-Kutta method \cite{Spalart:JCP1991} splits the governing equations
into their linear and non-linear components, \(\mathcal{L}\) and \(\mathcal{N}\) respectively.
Three sub-steps are performed to compute \(\mathbf{u}_{n+1}\) from \(\mathbf{u}_n\):
\be
    \label{eq:rk3_ss1}
    \mathbf{u}' = \mathbf{u}_n + \Delta t [\mathcal{L}(\alpha_1 \mathbf{u}_n + \beta_1 \mathbf{u}') + \gamma_1\mathcal{N}(\mathbf{u}_n)],
\ee
\be
    \label{eq:rk3_ss2}
    \mathbf{u}'' = \mathbf{u}' + \Delta t [\mathcal{L}(\alpha_2 \mathbf{u}' + \beta_2 \mathbf{u}'') + \gamma_2\mathcal{N}(\mathbf{u}') + \zeta_1\mathcal{N}(\mathbf{u}_n)],
\ee
\be
    \label{eq:rk3_ss3}
    \mathbf{u}_{n+1} = \mathbf{u}'' + \Delta t [\mathcal{L}(\alpha_3 \mathbf{u}'' + \beta_3 \mathbf{u}_{n+1}) + \gamma_3\mathcal{N}(\mathbf{u}'') + \zeta_2\mathcal{N}(\mathbf{u}')].
\ee
Here, \(\alpha_i\), \(\beta_i\), \(\gamma_i\) and \(\zeta_j\) for \(i = 1, 2, 3\) and \(j = 1, 2\) are parameters obtained from \citet{Spalart:JCP1991} and \citet{Orlandi:book}.
We split the non-linear term into two parts according to the skew symmetric form described in \citet{Morinishi:JCP1998}.
\be
    \mathbf{u}.\nabla\mathbf{u} = \frac{1}{2}\left(u_j\frac{\partial u_i}{\partial x_j}\right) + \frac{1}{2}\frac{\partial}{\partial x_j}(u_i u_j)
\ee
We use a predictor-corrector method at each sub-step of RK3.
Therefore, when we write (\ref{eq:nse_filt}) and (\ref{eq:temp_filt}) for the first sub-step of RK3 (Eq. \ref{eq:rk3_ss1}),
we first calculate the predicted values of velocity (\(\mathbf{u}^*\)) and temperature (\(T^*\)) as
\begin{equation*}
    \mathbf{u}^* = \mathbf{u}_n + \Delta t \left[\sqrt{\frac{Pr}{Ra}} \left(\alpha_1 \nabla^2 \mathbf{u}_n + \beta_1 \nabla^2 \mathbf{u}^*\right)
        + (\alpha_1 + \beta_1)\left(T_n \hat{z} - \nabla p_n\right) + \gamma_1\left(\nabla . \bm{\tau}_n - \mathbf{u}_n . \nabla \mathbf{u}_n\right)\right],
\end{equation*}
\be
    T^* = T_n + \Delta t \left[\sqrt{\frac{1}{Ra Pr}} \left(\alpha_1 \nabla^2 T_n + \beta_1 \nabla^2 T^*\right)
        + \gamma_1\left(\nabla . \mathbf{\sigma}_n - \mathbf{u}_n . \nabla T_n\right)\right].
\ee
We iteratively solve for \(\mathbf{u}^*\) and \(T^*\) using Jacobi iterative solver, and then compute the pressure correction, \(p_c\), as
\be
    \nabla^2 p_c = \frac{\nabla . \mathbf{u}^*}{(\alpha_1 + \beta_1)\Delta t}.
\ee
We solve the above equation using a geometric multigrid solver V-cycles.
The solver uses Red-Black Gauss-Seidel (RBGS) iterations for both smoothing at each multigrid level, as well as for solving the Poisson equation at the coarsest level.
The RBGS iterative solver was chosen for both fast convergence as well as for performing multithreaded vectorized calculations.
After obtaining \(p_c\), we then update pressure with the correction term, and correct the predicted velocity field to satisfy divergence:
\be
    p' = p_n + p_c, \qquad \qquad \mathbf{u}' = \mathbf{u}^* - \Delta t (\alpha_1 + \beta_1) \nabla p_c.
\ee
Finally we impose boundary conditions on \(\mathbf{u}'\), \(p'\) and \(T'\).
These steps are repeated for the two remaining sub-steps of RK3, (\ref{eq:rk3_ss2}) and (\ref{eq:rk3_ss3}), to finally obtain \(\mathbf{u}_{n+1}\), \(p_{n+1}\) and \(T_{n+1}\).


\section{Time series of global quantities}
\label{app:time_series}

\begin{figure}
    \centering
    \includegraphics[width=1.0\textwidth]{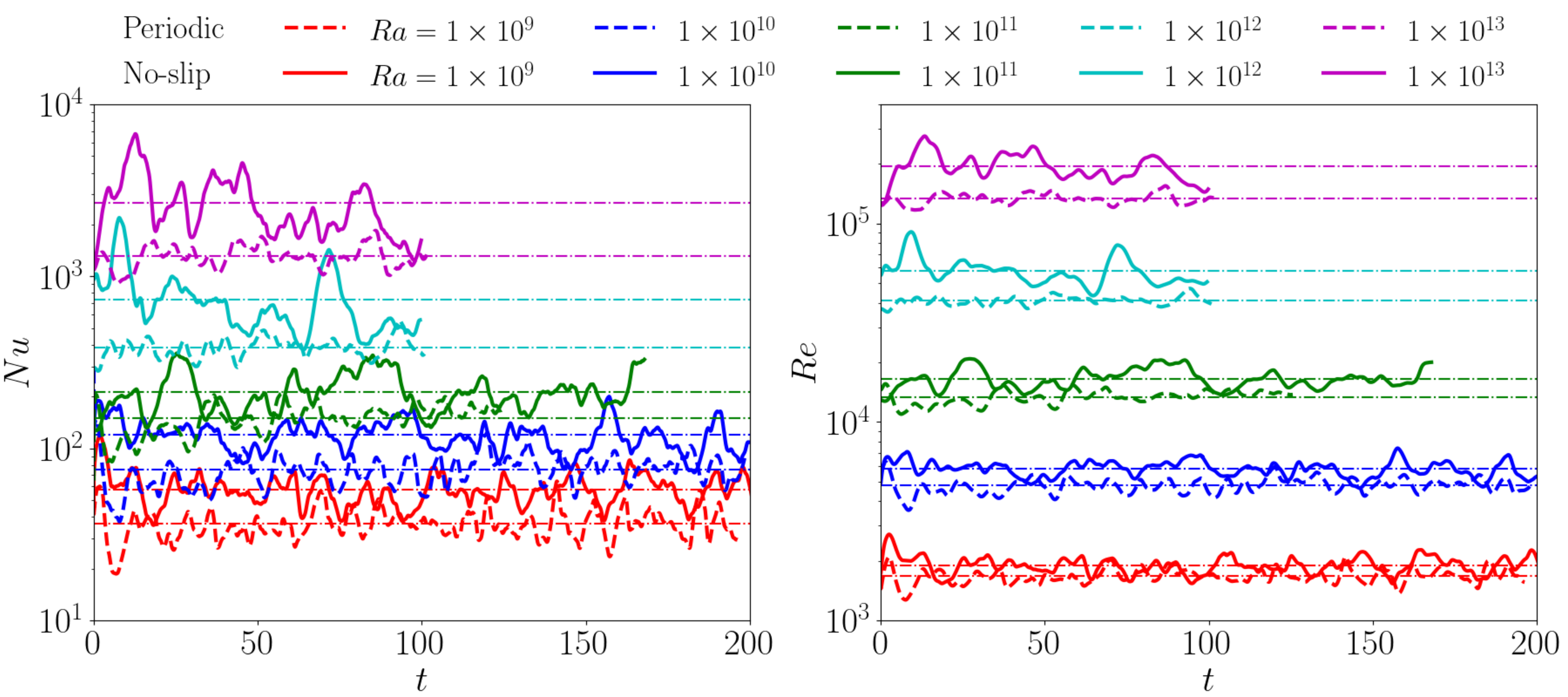}
    \caption{\label{fig:col_time_series}
        Time-series of Nusselt number (left frame) and Reynolds number (right frame) for thermal convection in a slender column as described in Section~\ref{sec:les_rbc} and \ref{sec:rbc_per}.
        The solid lines are for the case with adiabatic side-walls, whereas dashed lines represent the case with periodic side-walls.
        The time-scales of the fluctuations are such that the simulation has to be run for at least 100 units of free-fall time
        in order to obtain reliable estimates of time-averaged quantities.
            }
\end{figure}

We briefly look at the time-series of Nusselt number and Reynolds number for convection in a slender box of aspect ratio \(\Gamma = 1/10\),
described in Section~\ref{sec:les_rbc} and \ref{sec:rbc_per}.
Fig.~\ref{fig:col_time_series} shows the variation of \(Nu\) and \(Re\) in the left and right frames respectively for the two cases.
We limit the comparison here to the first five values of \(Ra\), from \(10^9\) to \(10^{13}\).
The results from the cases with adiabatic, no-slip walls are plotted with solid lines,
whereas the those from the cases with periodic boundary conditions in the horizontal directions are marked by dashed lines.
All plots are limited to a maximum non-dimensional time of \(t = 200\) in order to highlight the differences across the 5 decades of Rayleigh number.
As mentioned earlier in Section~\ref{sec:rbc_per}, we see that \(Re\) and \(Nu\) are reduced when we use periodic side-walls.
We attribute this to the lack of a large-scale structure of flow in the cavity.
Another conclusion to be made from the time-series data is that ideally, we need to perform the numerical experiments for reasonably long durations
to obtain reliable estimates of time-averaged quantities.
For all the cases up to \(Ra = 10^{13}\), we use a minimum of 100 free-fall units of time to calculate the averages.
Since direct simulations for such long time-spans can be computationally very exhaustive,
this is yet another persuasive argument for using LES to probe the physics of thermal convection at very high Rayleigh numbers.

\bibliography{bib/journal,bib/book}

\end{document}